\documentclass[%
 reprint, letterpaper,
showpacs,
preprintnumbers,
showkeys,
nofootinbib,
 amsmath,amssymb,
 aps, prd,
floatfix]{revtex4-1}
\usepackage{hyperref}
\usepackage[cp1250]{inputenc}
\newcommand{\ee}{\end{equation}}
\newcommand{\be}{\begin{equation}}
\newcommand{\ba}{\begin{array}}
\newcommand{\ea}{\end{array}}
\newcommand{\m}{M_{H^{\pm}}}
\newcommand{\g}{\ \mbox{GeV}}
\newcommand{\la}{\lambda_1}
\newcommand{\lb}{\lambda_2}
\newcommand{\lc}{\lambda_3}
\newcommand{\ld}{\lambda_4}
\newcommand{\lp}{\lambda_5}
\newcommand{\lcz}{\lambda_{45}}
\newcommand{\lczp}{\lambda_{345}}
\usepackage[pdftex]{graphicx}
\usepackage{amsmath}
\usepackage{amssymb}
\usepackage{url}
\usepackage{verbatim}
\usepackage{natbib}
\begin{document}
\title{Yukawa independent constraints for Two Higgs Doublet Models\\ with a 125 GeV Higgs boson}
\author{Bogumi\l a \'Swie\.zewska}
\email[e-mail address: ]{Bogumila.Swiezewska@fuw.edu.pl}
\affiliation{\textit{Faculty of Physics, University of Warsaw}\\ \textit{Ho\.za 69, 00-681 Warsaw, Poland}}
\pacs{12.60.Fr, 14.80.Ec, 14.80.Fd, 95.35.+d}
\keywords{Higgs boson, perturbative unitarity condition, oblique parameters, Two Higgs Doublet Models}
\preprint{IFT-6/2012}
\preprint{arXiv:1209.5725 [hep-ph]}

\setlength\arraycolsep{2pt}

\begin{abstract}

In this paper various  constraints  for the parameter spaces of two variants of the Two Higgs Doublet Model with a  $\mathbb{Z}_2$-symmetric potential are reconsidered, including the LHC data on existence of a $125\g$ Higgs-like boson. We analyze the model in which only one of the doublets develops a nonzero vacuum expectation value (VEV) -- the Inert Doublet Model (IDM) and the Mixed Model  where both of the doublets have nonzero VEVs. Positivity constraints, conditions determining the  type of the vacuum,  perturbative unitarity condition,  constraints following from electroweak precision tests together with the LEP bounds on masses of the scalars are included in the analysis. The analysis is performed without specific assumptions regarding the Yukawa sector. For the IDM constraints on quartic couplings and masses of the scalars as well as their implications for Dark Matter scenarios are presented. A new type of bound on the mass parameter of the potential coming from the condition for the existence of the Inert vacuum is given. In the Mixed Model a strong bound on the value of $\tan\beta$, $0.18\lesssim\tan\beta\lesssim5.59$, is found. It depends on the mass of the Higgs boson and is independent of the Yukawa interactions. Also Standard Model (SM)-like scenarios with either $h$ or $H$ playing the role of the SM-like Higgs boson are analyzed.
\end{abstract}

\maketitle

\section{Introduction}

When the most general $\mathbb{Z}_2$-symmetric potential of a  Two Higgs Doublet Model (2HDM) is assumed, still  different physical models can be realized, for a recent review see~\cite{Branco:2011}. Among these are the Inert Doublet Model (IDM)~\cite{Ma:1978, Barbieri:2006, Ma:2007} in which only one of the doublets develops nonzero vacuum expectation value (VEV) 
 and the Mixed Model where both of the doublets have nonzero VEVs. 

The aim of this work is to constrain the parameter spaces of these two 2HDMs in a consistent way, using the following: vacuum stability conditions,  perturbative unitarity condition, conditions determining the type of the vacuum (determining the validity regions of the Mixed Model and of the IDM), the electroweak precision tests (EWPTs) with the use of the $S$ and $T$ parameters and the LEP bounds on the scalars' masses.

In 2012 a Higgs-like particle of mass around $125\g$ was discovered at the LHC~\cite{atlas:2012, *cms:2012}. We assume that the only Higgs boson of the IDM corresponds to the discovered particle, while in the Mixed Model we consider two candidates which may play the role of the Standard Model (SM)-like Higgs boson.

Since there exist numerous distinct models of Yukawa interactions, this analysis does not involve model-dependent constraints from the Yukawa sector and is limited to the scalar sector, for the sake of clarity.

The perturbative unitarity condition was explored in the Mixed Model, following the approach of Refs.~\cite{Dicus:1973, Lee:1977prl, *Lee:1977}, by many authors, see Refs.~\cite{Weldon:1984, Huffel:1981,  *Casalbuoni:1986, *Casalbuoni:1988, *Maalampi:1991, Kanemura:1993, Akeroyd:2000, *Arhrib:2000, Ginzburg:2003, *Ginzburg:2005, Horejsi:2005}. Here we extend the analysis by considering the full tree-level high-energy scattering matrix and explicitly including the conditions for the existence of the Mixed vacuum. An extensive analysis of the parameter space of the 2HDM with two nonvanishing VEVs was performed in Ref.~\cite{Kanemura:2011}. However, this was done under certain assumptions, namely: soft  $\mathbb{Z}_2$ violation (fixed nonzero value of $m_{12}^2$), degeneracy of masses of $A$ and $H$ and the SM-like scenario. Here we consider a $\mathbb{Z}_2$-symmetric model ($m_{12}^2=0$); thus the results of \cite{Kanemura:2011} cannot be directly compared with ours. Another a\-na\-ly\-sis of the parameter space of 2HDM(II) is Ref.~\cite{Kang:2012}. The consequences of the stability and perturbativity conditions, that are assumed to be valid up to a certain cutoff scale, are analyzed there. Also the experimental data are incorporated. However, the main focus of~\cite{Kang:2012} is on the dependence of the allowed pa\-ra\-me\-ter space on the cutoff scale, which is different than in our case. In Ref.~\cite{Oneil:2012} oblique parameters in the most general $CP$-violating 2HDM (Mixed) were studied, their possible values  were discussed (also higher order parameters $V$, $W$ and $X$ were analyzed) and some bounds on the mass of the charged Higgs boson were found. We focus more on the impact of the EWPTs on the allowed regions of scalar masses in the 2HDM and on constraints for $\tan\beta$.

The unitarity constraints for the IDM were studied by us \cite{praca-mag, Gorczyca:2011} and were included also in analyses of Refs.~\cite{Arhrib:2012, Gustafsson:2012}. The EWPTs were analyzed for the IDM in Refs.~\cite{Barbieri:2006, Dolle:2009, Gustafsson:2012}. Reference~\cite{Gustafsson:2012} combines a wide range of constraints for the IDM, including various experimental results. However, there the main focus is on the possibility of accommodating in the IDM a heavy Higgs boson. Here, as was mentioned before, we focus on a $125\g$ Higgs boson.


The paper is organized as follows: in Sec.~\ref{sec:model} the~model is briefly introduced, the IDM and the Mixed Model are defined and constraints relevant for further analysis are presented. In Sec.~\ref{sec:unit} the standard perturbative unitarity approach as well as the oblique parameters are introduced and a short description of the method used to obtain the results is given. Next sections contain the results of the analysis: in Sec.~\ref{sec:param} bounds on the quartic parameters of the potential are given, and in Secs.~\ref{sec:inert} and~\ref{sec:mixed} constraints for the IDM and the Mixed Model are presented. Section~\ref{sec:sum} briefly summarizes the obtained results.

\section{The Models\label{sec:model}}


\subsection{Potential}

We consider a 2HDM with the following potential:
\be\label{pot}\renewcommand{\arraystretch}{1.5}
\begin{array}{rcl}
V&=&-\frac{1}{2}\left[m_{11}^{2}(\phi_{S}^{\dagger}\phi_{S})+m_{22}^{2}(\phi_{D}^{\dagger}\phi_{D})\right]\\*
&&+\frac{1}{2}\left[\lambda_{1}(\phi_{S}^{\dagger}\phi_{S})^{2}+\lambda_{2}(\phi_{D}^{\dagger}\phi_{D})^{2}\right]\\*
&&+\lambda_{3}(\phi_{S}^{\dagger}\phi_{S})(\phi_{D}^{\dagger}\phi_{D})+\lambda_{4}(\phi_{S}^{\dagger}\phi_{D})(\phi_{D}^{\dagger}\phi_{S})\\*
&&+\frac{1}{2}\lambda_{5}\left[(\phi_{S}^{\dagger}\phi_{D})^{2}+(\phi_{D}^{\dagger}\phi_{S})^{2}\right].\\*
\end{array}
\ee
The parameters $m_{11}^{2}$, $m_{22}^{2}$ and $\la,\ldots,\ld$ are real numbers and without loss of generality we can assume that $\lp<0$~\cite{Krawczyk:2010,  Krawczyk:2004sym, *Krawczyk:2004sym2, Branco:1999}.

Note that this potential is symmetric under two $\mathbb{Z}_2$ symmetries, the so-called $D$ symmetry  under action of which: $\phi_S\to\phi_S$ and  $\phi_D\to-\phi_D$  and $S$ symmetry which acts as follows: $\phi_S\to-\phi_S$ and $\phi_D\to\phi_D$~\cite{Krawczyk:2010}. $\mathbb{Z}_2$ symmetry prevents the existence of the  flavor changing neutral currents at the tree level. 

The requirement that the potential is bounded from below, which is necessary for a stable vacuum state to exist, leads to the following conditions (vacuum stability conditions/positivity constraints)~\cite{Ma:1978}:
\be\label{pos}
\la>0,\quad\lb>0,\quad\lc+\sqrt{\la\lb}>0,\quad\lczp+\sqrt{\la\lb}>0,
\ee
where $\lczp=\lc+\ld+\lp$.

A state of the lowest energy, for which the potential approaches its global minimum, is the vacuum state. The potential (\ref{pot}) can develop five different types of mi\-ni\-ma. It depends on the values of parameters, which of the minima is the global one. Models in which scalars interact according to the potential (\ref{pot}) but distinct vacua are realized, differ significantly in physical content as they develop different particle spectra. Here we consider two of them,\footnote{In the remaining ones unphysical effects, like massive photons in a model with charge breaking vacuum or massless fermions in the Inert-like model, are present \cite{Krawczyk:2010}.} to be presented below.


\subsection{The Inert Doublet Model}

The IDM is defined as a 2HDM with the potential~(\ref{pot}), an Inert vacuum state and Model~I of Yukawa interactions \cite{Ma:1978, Barbieri:2006, Ma:2007}. According to the Model~I, only $\phi_S$ couples to fermions. We fix a model of Yukawa interactions for completeness, and our results are independent of Yukawa sector [only for the dark matter (DM) considerations do interactions with fermions have to be defined].

The Inert vacuum is realized when only the $\phi_S$ doublet has a non-vanishing VEV. We take \renewcommand{\arraystretch}{.7}$\langle\phi_{S}^0\rangle=\frac{v}{\sqrt{2}}$, $\langle\phi_{D}^0\rangle=0$, $v=246\g$. The doublets can be decomposed as follows
$$
\phi_{S}=\left(\begin{array}{c}\renewcommand{\arraystretch}{1.3}
G^{+}\\
\frac{1}{\sqrt{2}}(v+h+iG)\end{array}\right),\quad
\phi_{D}=\left(\begin{array}{c}\renewcommand{\arraystretch}{1.3}
H^{+}\\
\frac{1}{\sqrt{2}}(H+iA)\end{array}\right).\nonumber
$$
In this case the matrix of the second derivatives of the potential is diagonal, so the fields $G^{\pm}$, $H^{\pm}$, $h$, $G$, $H$, $A$ are mass eigenstates. The particles having their origin in the $\phi_D$ doublet ($H^{\pm}$, $H$, $A$) are called the dark or inert scalars. Their masses read:\renewcommand{\arraystretch}{1.5}
\be\label{masyinert}
\begin{array}{rcl}
M_{h}^{2}&=&m_{11}^{2}=\la v^{2},\\*
M_{H^{\pm}}^{2}&=&\frac{1}{2}(\lc v^{2}-m_{22}^{2}),\\*
M_{A}^{2}&=&\frac{1}{2}(\lczp^{-}v^{2}-m_{22}^{2})=\m^2+\frac{1}{2}\lambda_{45}^- v^2,\\*
M_{H}^{2}&=&\frac{1}{2}(\lczp v^{2}-m_{22}^{2})=\m^2+\frac{1}{2}\lambda_{45} v^2,\\*
\ea
\ee
where $\lczp^-=\lc+\ld-\lp,\; \lambda_{45}^-=\ld-\lp,\;\lambda_{45}=\ld+\lp $. $G^{\pm}$ and $G$ are massless (would-be) Goldstone bosons. 

The Inert vacuum can be realized  only if the following conditions are fulfilled~\cite{Krawczyk:2010, praca-mag}:
\be\label{inert-war}
M_h^2,\, M_H^2,\,M_A^2,\,\m^2\geqslant0,\quad\frac{m_{11}^2}{\sqrt{\la}}>\frac{m_{22}^2}{\sqrt{\lb}},
\ee
where the masses of the particles are defined by Eq.\,(\ref{masyinert}). Positivity of the masses squared guarantees that the Inert state is the minimum of the potential. The other condition assures that the Inert minimum is a global one, having lower energy than the coexisting Inert-like minimum.

In the IDM the unique Higgs boson is the $h$ particle. It couples (at the tree level) to gauge bosons and fermions just like the SM Higgs. Thus, $h$ is SM-like and we assume that it corresponds to the boson discovered at the LHC~\cite{atlas:2012, cms:2012} and set $M_h=125\g$.

As the Inert vacuum preserves the $D$ symmetry and so does the Lagrangian, this symmetry is exact in the IDM. Therefore the lightest $D$-odd particle is stable and can be a good DM candidate given that it is electrically neutral. Here we assume that $H$ is the DM candidate \cite{Krawczyk:2010}, so the condition $\ld+\lp<0$ must hold\footnote{This condition holds also in the Mixed Model, see Eq.\,(\ref{mixed-war}). Then, the last two positivity conditions in Eq.\,(\ref{pos}), reduce to the last one.}; see Eq.~(\ref{masyinert}).

We impose the LEP bounds on the masses of the scalars~\cite{Gustafsson:2009,*Gustafsson:2010}, namely
\be
\begin{array}{c}
\m +M_H>M_{W},\ \m+ M_A>M_W,\\*
M_H+M_A >M_Z,\ \m>70\g \\*
\textrm{and exclude the region where simultaneously:} \\*
M_H< 80\g,\ M_A< 100\g,\ 
M_A - M_H> 8\g.\label{lep-inert}\end{array}\ee


\subsection{The Mixed Model}

The Mixed Model is defined as a 2HDM with a potential described by Eq.~(\ref{pot}), vacuum state of a Mixed form and  Model II of Yukawa interactions. We specify the form of interaction with fermions only for completeness as it does not affect our analysis. In particular all the results obtained in this analysis are valid also for the Model~I of Yukawa interactions.

The Mixed vacuum state is realized when both of the doublets develop nonzero VEVs: \renewcommand{\arraystretch}{.7}
$\langle\phi_{S}^0\rangle=\frac{v_S}{\sqrt{2}}$, $\langle\phi_{D}^0\rangle=\frac{v_D}{\sqrt{2}}$, $v_{S}$, $v_{D}\neq0$,  $v^{2}=v_{S}^{2}+v_{D}^{2}$. The scalar fields can be  represented as follows:\renewcommand{\arraystretch}{1.3}
\begin{eqnarray}
\phi_{S}&=&\left(\begin{array}{c}
\rho_{S}^{+}\\
\frac{{1}}{\sqrt{2}}(v_{S}+\rho_{S}+i\chi_{S})\end{array}\right),\nonumber\\
\quad\phi_{D}&=&\left(\begin{array}{c}
\rho_{D}^{+}\\
\frac{1}{\sqrt{2}}(v_{D}+\rho_{D}+i\chi_{D})\end{array}\right).\nonumber
\end{eqnarray}
Mass eigenstates are mixtures of the $\rho_{K}^{\pm}$, $\rho_{K}$ and $\chi_{K}$ (see for example~\cite{Krawczyk:2010,Krawczyk:2004sym, Krawczyk:2004sym2}), namely\renewcommand{\arraystretch}{1}
$$
\begin{array}{c}
\left(\begin{array}{c}
G^{\pm}\\
H^{\pm}\\
\end{array}\right)=
R(\beta)\left(\begin{array}{c}
\rho_S^{\pm}\\
\rho_D^{\pm}\\
\end{array}\right), \quad
\left(\begin{array}{c}
G\\
A\\
\end{array}\right)=
R(\beta)\left(\begin{array}{c}
\chi_S\\
\chi_D\\
\end{array}\right), \\*[8pt]
\left(\begin{array}{c}
H\\
h\\
\end{array}\right)=
R(\alpha)\left(\begin{array}{c}
\rho_S\\
\rho_D\\
\end{array}\right),
\end{array}
$$
where $R$ denotes a rotation matrix of an angle  $\alpha$ or $\beta$, $\alpha\in(-\pi/2,\,\pi/2)$, $\beta\in(0,\,\pi/2)$ and $\tan\beta=\frac{v_D}{v_S}$.  
 $G^{\pm}$ and $G$ stand for the Goldstone bosons and $H^{\pm}$, $A$, $H$, $h$ are the physical particles of masses:\renewcommand{\arraystretch}{1.5}
\be\label{masymix}
\begin{split}
M_{H^{\pm}}^{2}=&-\frac{v^2}{2}(\ld+\lp),\\*
M_{A}^{2}=&-\lp v^{2},\\*
M_{H}^{2}=&\frac{v^{2}}{2}\frac{1}{1+\tan^{2}\beta}\Big(\la+\lb\tan^{2}\beta\\*
&+\sqrt{(\la-\lb\tan^{2}\beta)^{2}+4\lambda_{345}^{2}\tan^{2}\beta}\Big),\\*
M_{h}^{2}=&\frac{v^{2}}{2}\frac{1}{1+\tan^{2}\beta}\Big(\la+\lb\tan^{2}\beta\\*
&-\sqrt{(\la-\lb\tan^{2}\beta)^{2}+4\lambda_{345}^{2}\tan^{2}\beta}\Big).\\*
\end{split}
\ee

The relative (with respect to SM) strength of the coupling of $h$ and $H$ to the gauge bosons is controlled by  $\beta-\alpha$. Namely, $\sin(\beta-\alpha)$ corresponds to the  $hW^+W^-$ and $hZZ$ vertices and $\cos(\beta-\alpha)$ to $HW^+W^-$ and $HZZ$ vertices. The role of a SM-like Higgs particle can be played either by $h$ [when $\sin(\beta-\alpha)=1$] or by $H$ [when $\cos(\beta-\alpha)=1$].

The  global minimum of the potential is of the Mixed type (i.e., Mixed vacuum is realized)  only if the following conditions are sa\-tis\-fied~\cite{Krawczyk:2010, praca-mag}:
\be\label{mixed-war}
\begin{split}
&v_{S}^{2}=\frac{m_{11}^{2}\lb-\lczp m_{22}^{2}}{\la\lb-\lczp^{2}}>0,\\
&v_{D}^{2}=\frac{m_{22}^{2}\la-\lczp m_{11}^{2}}{\la\lb-\lczp^{2}}>0,\\[.2pt]
&\ld+\lp<0,\quad \lp<0,\quad \la\lb-\lczp^{2}>0.
\end{split}
\ee

In the Mixed Model we adopt the LEP bound on the mass of the charged Higgs boson~\cite{Pierce:2007, Branco:2011}:
\be
\m>78\g,\label{lep-mixed}
\ee
which is valid for different types of Yukawa interactions.\footnote{The combined LEP results~\cite{Abbiendi:2013} give slightly weaker bound for the Model I, namely $\m>72.5\g$ for $M_A>12\g$ but it does not affect our results significantly.}


\section{Method of the analysis\label{sec:unit}}


\subsection{Unitarity constraints\label{sec:warunit}}

Analyzing the unitarity constraints we followed the standard approach of Refs. \cite{Lee:1977prl, Lee:1977, Dicus:1973, Huffel:1981, Casalbuoni:1986, Casalbuoni:1988, Maalampi:1991, Kanemura:1993, Akeroyd:2000, Arhrib:2000, Ginzburg:2003, Ginzburg:2005, Horejsi:2005}. From unitarity of the $S$ matrix 
 it follows that the full partial wave amplitudes should lie on the so-called Argand circle. In particular, in the limit of momentum transfer much greater than the masses of the particles involved in the scattering, inequality $|\Re(a^{(j)}(s))|\leqslant\frac{1}{2}$ holds for elastic scattering partial wave amplitudes $a^{(j)}$. For the theory to be perturbative, the zeroth order amplitudes should not lie too far from the circle, which can be assured by assuming that~\cite{Kanemura:1993, Akeroyd:2000, Arhrib:2000, Horejsi:2005}:
\be\label{warunit}
|\Re(a_{0}^{(0)}(s))|\leqslant\frac{1}{2},
\ee
where $a_{0}^{(0)}(s)$ denotes the tree-level amplitude of the $s$ wave.\footnote{The choice of the perturbative unitarity condition is to some extent arbitrary; the following form $|a_{0}^{(0)}|\leqslant1$ was also used in the literature~\cite{Lee:1977prl, Lee:1977, Huffel:1981, Weldon:1984, Casalbuoni:1986, Casalbuoni:1988, Maalampi:1991, Ginzburg:2003, Ginzburg:2005}, leading to differences between results.}

It is well known that a possible threat to unitarity is due to the scattering of the longitudinally polarized vector bosons. The scalar particles are supposed to unitarize the scattering amplitudes. Because of the equivalence theorem \cite{Cornwall:1974, Lee:1977prl, Lee:1977}, in the high-energy limit it is sufficient to take into account scattering of the Goldstone bosons instead of the longitudinally polarized vector bosons. In addition, in the same regime, the terms corresponding to cubic couplings are suppressed and only the quartic terms are relevant. Moreover, as the physical fields (mass eigenstates) are obtained from the original fields appearing in the basic Lagrangian by means of a unitary transformation and only the eigenvalues of the scattering matrix are important for the following analysis, it suffices to consider the scattering matrix between the original fields~\cite{Kanemura:1993}.  So, using the high-energy formula for $a_0^{(0)}(s)$, the unitarity condition, Eq.~(\ref{warunit}), can be reexpressed in terms of the eigenvalues $\Lambda_i$ of the scattering matrix: 
\be\label{warunit2}
|\Lambda_i|\leqslant8\pi.
\ee

We considered the \emph{full} tree-level high-energy scattering matrix (of dimension 25) of the scalar sector. There are 14 neutral channels~\cite{Kanemura:1993}, eight charged channels~\cite{Akeroyd:2000, Arhrib:2000} and three doubly charged channels~\cite{Ginzburg:2003, Ginzburg:2005, praca-mag, Gorczyca:2011}.\footnote{In Ref.~\cite{Ginzburg:2003, Ginzburg:2005} different classification of states was used, and the doubly charged states were contained in states of hypercharge 2.}  Dia\-go\-na\-li\-za\-tion of the full scattering matrix leads to 12 distinct eigenvalues being functions of  the parameters $\lambda_i$. 
Applying the perturbative unitarity condition (\ref{warunit2}) to these eigenvalues yields a set of 12 inequalities for $\lambda_i$. 

Equivalently,  different set of parameters can be chosen. The IDM can be parametrized by the parameters $\lambda_i$ and $m_{22}^2$ [$m_{11}^2$ is just a function of $\la$, Eq.~(\ref{masyinert})] or, for example, by the masses of the scalar particles, $\lb$ and $m_{22}^2$.  The Mixed Model can be described by $\lambda_i$ and $\tan\beta$ or, equivalently, by the masses of the scalars, $\tan\beta$ and $\sin\alpha$ [or $\sin(\beta-\alpha$)]. Rewriting the inequalities following from the unitarity condition in terms of the latter sets of parameters gives bounds on the masses of the scalars and, possibly, on the remaining parameters.


\subsection{Oblique parameters}

\subsubsection{Definition of $S$ and $T$}
With the use of the oblique parameters the contributions from the new physics (NP) to the electroweak processes can be easily tracked. We followed the definitions  from \cite{Grimus:2008, *Grimus:2008t} (see references therein). 
 Namely
\begin{align}\setlength\arraycolsep{2pt}
T&=\frac{1}{\alpha}\bigg(\frac{A_{WW}(0)}{M_W^2} - \frac{A_{ZZ}(0)}{M_Z^2}\bigg),\nonumber\\*[4pt]
S&=\frac{4 s_W^2 c_W^2}{\alpha}\bigg(\frac{A_{ZZ}(M_Z^2)-A_{ZZ}(0)}{M_Z^2}\\*
&\phantom{=;}-\frac{\partial A_{\gamma\gamma}(q^2)}{\partial q^2}\bigg\arrowvert_{q^2=0}+\frac{c_W^2-s_W^2}{c_W s_W}\frac{\partial A_{\gamma Z}(q^2)}{\partial q^2}\bigg\arrowvert_{q^2=0}\bigg),\nonumber
\end{align}
where $\alpha=e^2/(4\pi)$ is the fine-structure constant, $s_W=\sin\theta_W$, $c_W=\cos\theta_W$ are the sine and cosine, respectively, of weak mixing angle and $A_{VV'}$ is defined as follows:
\begin{displaymath}
\Pi_{VV'}^{\mu\nu}=g^{\mu\nu}A_{VV'}(q^2)+q^{\mu}q^{\nu}B_{VV'}(q^2).
\end{displaymath}
Here $\Pi_{VV'}^{\mu\nu}$ is the vacuum-polarization tensor and $VV'$ denotes the divectors: $\gamma\gamma$, $\gamma Z$, $ZZ$ or $WW$. Moreover, one has to remember that $A_{VV'}(q^2)$ contains only contributions from the NP, namely
$$
A_{VV'}(q^2)=A^{\textrm{full}}_{VV'}(q^2)-A^{\textrm{SM}}_{VV'}(q^2),
$$
where $A^{\textrm{full}}_{VV'}(q^2)$ denotes the quantity calculated in considered model (in this case 2HDM) and $A^{\textrm{SM}}_{VV'}(q^2)$ denotes the same quantity computed in the SM.

The formulas used to compute the values of the $S$ and $T$ parameters can be found in the Appendix.

\subsubsection{Values of $S$ and $T$\label{st-values}}
The $S$ and $T$ parameters were found to have the following values~\cite{Nakamura:2010}\footnote{The values of $S$ and $T$ have been updated recently. The new values read $S=0.04\pm0.09,\ T=0.07\pm0.08$, with correlation $\rho=88\%$; see~\cite{Beringer:2012}. However, the change hardly influences our results.} (for $U$ fixed to be equal~0):
\begin{equation}\label{st-fit}
\begin{split}
S&=0.03\pm0.09,\\
T&=0.07\pm0.08.
\end{split}
\end{equation}
The correlation between $S$ and $T$ is equal to 87\%. Using the program \cite{STelipse} modified by us we generated the $1\sigma$ and $2\sigma$ ellipses. We assumed that a point  fulfills the experimental requirements if it falls inside the $2\sigma$ ellipse in the $(S,T)$ space.


\subsection{Method of the analysis}

We scanned randomly the parameter spaces of the IDM and the Mixed Model checking whether the following conditions were fulfilled:
\begin{subequations}\label{warunki}
\begin{equation}\label{war}
\begin{split}
&\textrm{- positivity constraints, Eq.\,(\ref{pos})};\\*
&\textrm{- conditions determining the type of vacuum:}\\*
&\phantom{- }\textrm{Eq.\,(\ref{inert-war}) or Eq.\,(\ref{mixed-war}) respectively};\\*
&\textrm{- perturbative unitarity condition, Eq.\,(\ref{warunit2})};\\*
&\textrm{- }\lp<0\;\textrm{and}\; \ld+\lp<0;\\*
&\textrm{- the LEP bounds, Eq.~(\ref{lep-inert}) or~Eq.~(\ref{lep-mixed}).}
\end{split}
\end{equation}
To constrain the masses of the scalar particles we took into account also:
\begin{equation}\label{ewpt}
\begin{split}
&\textrm{- EWPTs ($2\sigma$)---see Sec.~\ref{st-values}. $\phantom{\textrm{or    aEq.\,(\ref{inert-war})}}$}
\end{split}
\end{equation}
\end{subequations}
Separate scans were performed for the case when the model is parametrized by the parameters $\lambda_i$ and when it is parametrized by the masses of the scalars.

In the numerical analysis the values of the parameters $\lambda_i$ were chosen randomly from the following ranges: $\la,\,\lb\in(0,\,35]$, $\lp\in[-20,\,0)$, $\ld\in[-30,-\lp)$, $\lc\in(-\sqrt{\la\lb}-\ld-\lp,\,35]$.

In the IDM the mass of $h$ was set to  $M_h=125\g$. Remaining masses were in the ranges: $M_A,\,\m\in(0,\,1010]\g$, $M_H\in(0,\,\min(M_A,\,\m)]$, because $H$ is supposed to be the DM candidate and thus has to be the lightest of the dark scalars. $\lb$ is in principle allowed to be in the region $(0,\,35]$; however it is also constrained by Eqs.\,(\ref{pos}) and (\ref{inert-war}).

In the Mixed Model the masses are allowed to be in the following ranges $\m,\,M_H,\,M_A\in(0,800]\g$, $M_h\in(0,\,M_H]$ [see Eq.\,(\ref{masymix})], $\tan\beta\in [0,\,60]$\footnote{There exists a lower bound on the value of $\tan\beta$ ($\tan\beta\geqslant0.29$) which comes from the assumption of perturbativity of the $\overline{t}bH^{\pm}$ coupling and is valid in Models I-IV of Yukawa interactions~\cite{Barger:1989}. We did not impose it here directly in order to observe the impact of the considered conditions on the small values of $\tan\beta$.} and $\sin\alpha\in[-1,\,1]$. 

As a result of the scans we present overall numerical bounds on the parameters of interest and scatter plots showing the allowed regions in the parameter space. The figures displayed are projections of multidimensional regions onto two-dimensional planes of selected parameters. Normally (if not stated differently in the caption of a figure) the dark green (gray) regions correspond to the outcome of a scan with conditions~(\ref{war}) imposed, while the light green (gray) regions plotted on the top of the dark ones are what remains after imposing the full set of conditions~(\ref{warunki}): (\ref{war}) and~(\ref{ewpt}). Therefore the light green (gray) domains are always subsets of the dark regions. With more sample points, the boundaries of the regions get sharper.

\section{Constraints on quartic parameters of the potential\label{sec:param}}

In this section we present constraints on the parameters $\lambda_i$.  These parameters can be treated in two different ways.

First of all, $\lambda_i$ may be simply treated as the parameters describing the potential~(\ref{pot}), with no reference to a particular vacuum state. In this approach $\lambda_i$ can be constrained with the use of the conditions~(\ref{war})  (of course without the conditions determining the type of the vacuum and the LEP bounds). Resulting viable regions in the $\lambda_i$ parameter space are then valid for any model built upon the potential~(\ref{pot}), regardless of the type of the vacuum chosen. Examples of such regions in $(\lc,\ld)$ and $(\lb,\lczp)$ planes, are presented in Figs.~\ref{fig:l1} and~\ref{fig:l2} in dark green (gray).  One can see that the parameters considered here are  correlated, whereas a corresponding plot in $(\la,\lb)$ space is a rectangle. The numerical upper and lower bounds for $\lambda_i$ are shown below in Eq.~(\ref{wyniki lambdy}) (values without brackets).

Origin of some of the bounds visible in the plots can be easily identified, for example the sharp cutoff in the lower right part of the plot in Figs.~\ref{fig:l1} and~\ref{fig:l2} [upper panel, dark green (gray)] corresponds to the unitarity bound on one of the eigenvalues of the scattering matrix: $|\lc-\ld|\leqslant 8\pi$, which implies $\ld\geqslant\lc-8\pi$ for $\lc\geqslant\ld$. Similarly the upper bound on $\lb$ visible in Figs.~\ref{fig:l1} and~\ref{fig:l2} [lower panel, dark green (gray)] originates from constraint on another eigenvalue: $\left|\frac{1}{2}\left(3\left(\la+\lb\right)+\sqrt{9\left(\la-\lb\right)^2+4(2\lc+\ld)^2}\right)\right|\leqslant 8\pi$, which for $\la=\lb$, $2\lc=-\ld$ gives $\lb\leqslant \frac{8\pi}{3}\approx 8.38$. The lower border of the region in Fig.~\ref{fig:l1} corresponds to the positivity constraint, Eq.~(\ref{pos}), $\lczp+\sqrt{\la\lb}>0$ with maximal value of $\la$ inserted [see Eq.~(\ref{wyniki lambdy})]: $\la=8.38$ for the dark green (gray) region and $\la=0.26$ for the light green (gray). It is thus visible that the upper bounds presented in Eq.~(\ref{wyniki lambdy}) depend on the choice of the perturbative unitarity criterion,~Eq.~(\ref{warunit}). Choosing a more conservative one would result in more stringent limits on $\lambda_i$, whereas relaxing~Eq.~(\ref{warunit}) would allow for larger $\lambda_i$.
\begin{figure}[h]
\centering
\includegraphics[width=\columnwidth]{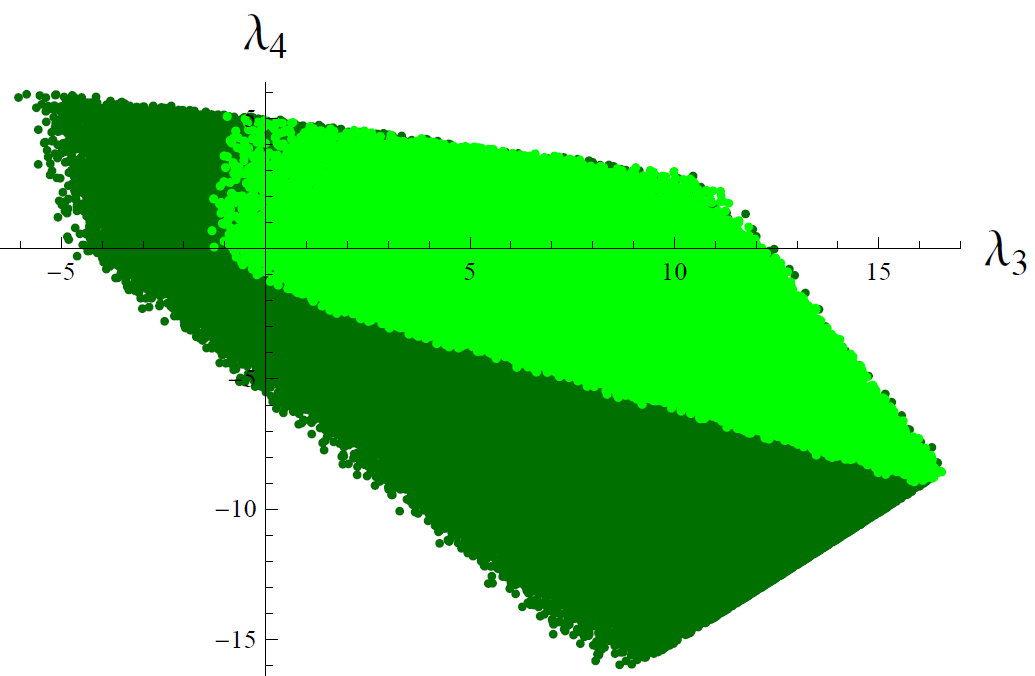}
\includegraphics[width=\columnwidth]{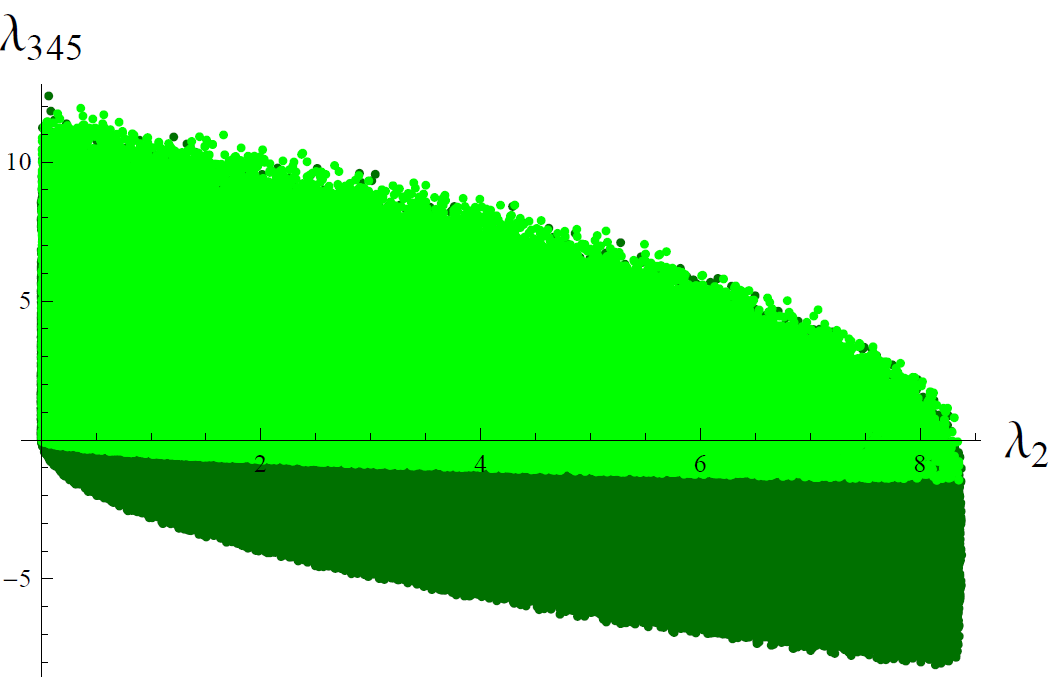}
\caption{Comparison of the regions of the parameter space allowed  by the constraints (\ref{war}) (without conditions determining type of the vacuum and the LEP bounds) in any model built upon potential~(\ref{pot}) [dark green (gray)] with the regions allowed by the full set of conditions~(\ref{war}),~(\ref{ewpt}) in the IDM  [light green (gray)]. The light green (gray) region is plotted on the top of the dark green (gray) and is a subset of it.
\label{fig:l1}}
\end{figure}
\begin{figure}[h]
\centering
\includegraphics[width=\columnwidth]{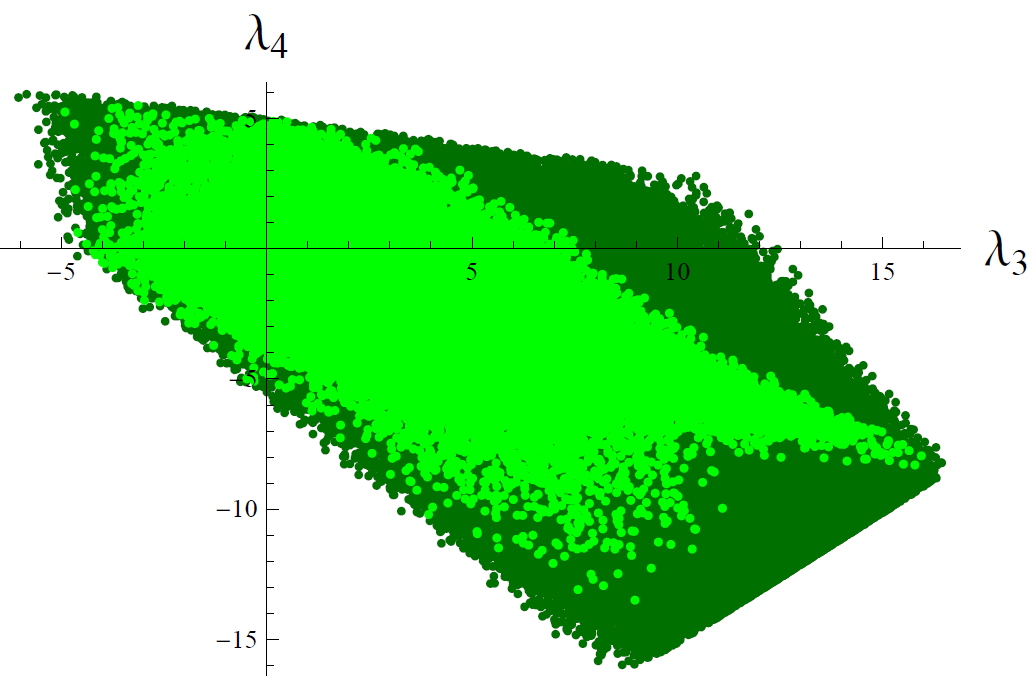}
\includegraphics[width=\columnwidth]{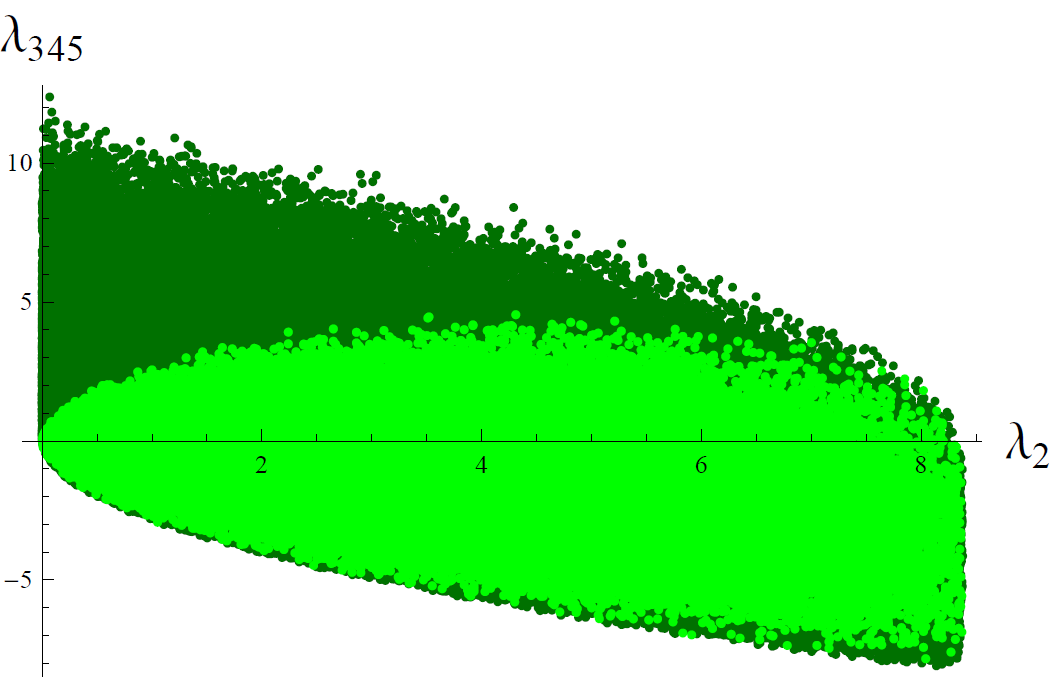}
\caption{Comparison of the regions of the parameter space allowed  by the constraints (\ref{war}) (without conditions determining type of the vacuum and the LEP bounds) in any model built upon potential~(\ref{pot}) [dark green (gray)] with the regions allowed by the full set of conditions~(\ref{war}),~(\ref{ewpt}) in the Mixed Model [light green (gray)]. The light green (gray) region is plotted on the top of the dark green (gray) and is a subset of it.
\label{fig:l2}}
\end{figure}

Moreover, we can also constrain $\lambda_i$ within particular models: the IDM or the Mixed Model. Then the full set of conditions~(\ref{warunki}) should be applied. The resulting allowed regions are shown in light green (gray) in Figs.~\ref{fig:l1} (for the IDM with $M_h=125\g$) and~\ref{fig:l2} (for the Mixed Model), for comparison plotted on the top of the more general dark green (gray) regions discussed above.  The viable regions for $\lambda_i$ in the IDM strongly differ from the ones valid in the Mixed Model. The differences in the results obtained for the two models show the importance of the conditions determining the type of the vacuum. The numerical bounds on $\lambda_i$ are presented below (in the round brackets for the IDM and in the square brackets for the Mixed Model):
\renewcommand{\arraystretch}{1.1}
\be\label{wyniki lambdy}
\begin{array}{r@{\;\leqslant\;}c@{\;\leqslant\;}lcccc}
\left[0\right] \ (0.26) \ 0&\la&8.38\ (0.26) \  [8.38],\\*
\left[0\right] \  (0) \  0&\lb&8.38\  (8.38) \  \left[8.38\right],\\*
\left[-4.92\right]\ (-1.32)\ -6.05&\lc&16.53\ (16.53)\ [15.95],\\*
\left[-13.49\right]\ (-8.95)\ -15.98&\ld&5.93\ (5.08)\ [5.50],\\*
\left[-8.06\right]\ (-8.22)\ -8.34&\lp&0\ (0)\ [0].\\*
\end{array}
\ee

As the conditions~(\ref{war}) include perturbative unitarity condition, the inequalities above show the region of perturbativity of the theory. Most of the bounds are significantly more stringent than the frequently used bound: $|\lambda_i|<4\pi$.


\section{Results for the IDM\label{sec:inert}}


\subsection{Constraints on the quartic couplings}

In the IDM the quartic coupling constants between physical particles are simple combinations of the parameters $\lambda_i$, for example $\lb$ represents  the $H^+H^-HH$ coupling and $\lczp$ the $hhHH$ coupling. $\lczp$ also controls the annihilation $HH\to h$, so the DM relic density is sensitive to its value.  The quartic couplings can be constrained with the use of the conditions~(\ref{warunki}), just like $\lambda_i$ in the previous section ($\lambda_{ij}=\lambda_i+\lambda_j$). This gives the following bounds:
\renewcommand{\arraystretch}{1.15}
\be\label{wynik sprz}
\begin{array}{r@{\;\leqslant\;}c@{\;\leqslant\;}lcccc}
-1.45&\lczp&11.94,\\*
-1.15&\lczp^{-}&16.40,\\*
-8.33&\frac{1}{2}\lcz&0,\\*
-2.64&\frac{1}{2}\lcz^{-}&5.08,\\*
-1.22&\lambda_{34}&13.34\\*
\end{array}
\ee
\renewcommand{\arraystretch}{1}
The correlation between $\lczp$ and $\lb$ in the IDM is presented in Fig.~\ref{fig:l1} (lower panel). 
The correlations between parameters may be source of additional constraints. For  example, it has been shown recently in Ref.~\cite{Arhrib:2012} that the enhancement in the $h\to\gamma\gamma$ channel in the IDM is possible only for $\lambda_3<0$ ($\lc$ is the coupling between $h$ and $H^+H^-$). So, as follows from Fig.~\ref{fig:l1} (upper panel), the enhancement in the $h\to\gamma\gamma$ channel is only possible for $-1.5\lesssim\ld\lesssim5$.

Of course the results for the IDM presented in Eqs.~(\ref{wyniki lambdy}) and~(\ref{wynik sprz}) depend on the mass of the Higgs boson, here $M_h=125\g$. If $M_h=126\g$, maximal shifts in the results would be at the level of 0.01, so basically unobservable because of the accuracy of the numerical method that was applied. If a heavier Higgs boson was considered, e.g. with  mass $M_h=200\g$, the bounds would shift more significantly, starting from $\lambda_1=(\frac{M_h}{v})^2\approx 0.66$ and the parameters related to it via positivity constraints, i.e., $\lc,\ \lczp,\ \lczp^-$ ($\lczp^->\lczp$ because $\lp<0$), lower bounds on which would be around $-2.3$. Other bounds would be only slightly changed, with relative shift at the level of a few percent.


\subsection{Constraints on masses}

In the IDM $h$ plays the role of the SM Higgs boson so we fixed its mass to  $M_h=125\g$ [so $m_{11}^2=(125\g)^2$ and $\la=0.26$]. The parameter $m_{22}^2$ enters the expressions for the masses of the dark scalars [Eq.~(\ref{masyinert})], but does not enter the scattering matrix (Sec.~\ref{sec:warunit}). Thus, the bounds on masses depend on its value.  For the case with $m_{22}^2=0$ the bounds on the masses of scalars following from the  constraints~(\ref{war}) read
\be\label{inertwyn}
\begin{array}{r@{\;\leqslant\;}l}
M_{H}&602\,\textrm{GeV},\\*
M_{H^{\pm}}&708\,\textrm{GeV},\\*
M_{A}&708\,\textrm{GeV.}\\*
\end{array}
\ee
These constraints are mainly due to the perturbative unitarity condition.

We have checked that in a large range of $m_{22}^2$, for $|m_{22}^2|\lesssim10^4\g^2$,   the results hardly change with respect to the case with $m_{22}^2=0$. However, when $m_{22}^2$ is extremely negative, the allowed regions of masses are changed, also lower bounds on masses develop. The regions of dark scalars' masses allowed by the conditions (\ref{warunki}) for both of the cases are presented in Fig.~\ref{fig:inert} in the $(\m,M_A)$ and $(\m,M_H)$ planes. The graph representing correlations in the  $(M_A,M_H)$ plane is not displayed as it is very similar to the one for $(\m,M_H)$. Note that the regions of masses allowed for the cases with $m_{22}^2=0$ and $m_{22}^2=-10^6\g^2$ have empty intersection (see Fig.~\ref{fig:inert}, lower panel).

\begin{figure}[ht]
\centering
\includegraphics[width=\columnwidth]{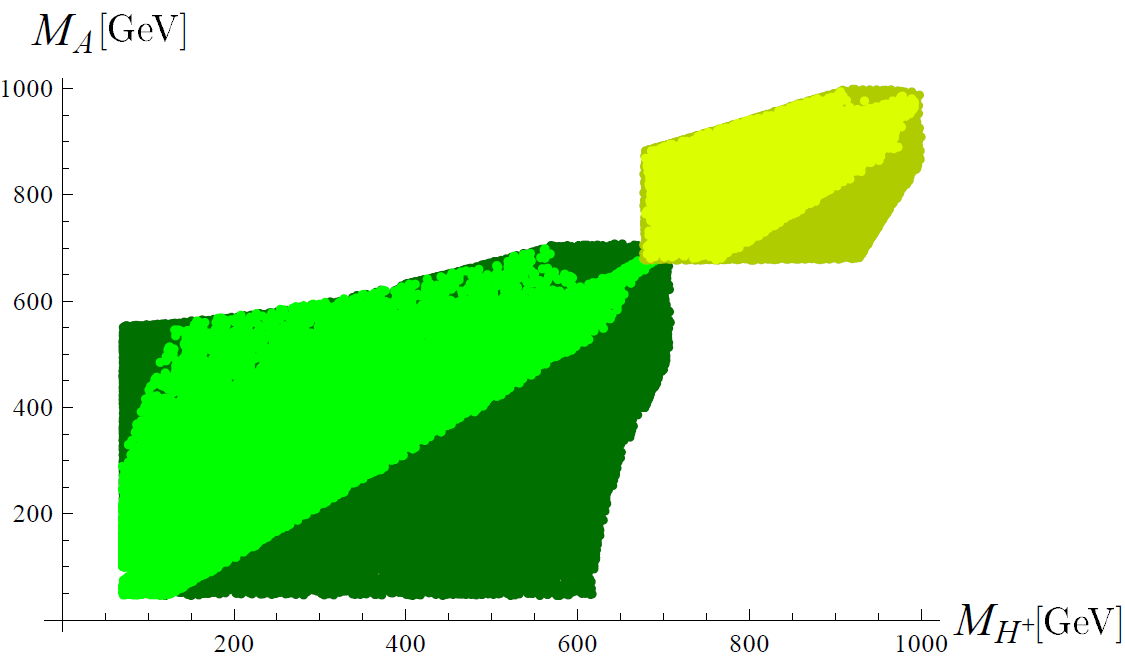}\\
\includegraphics[width=\columnwidth]{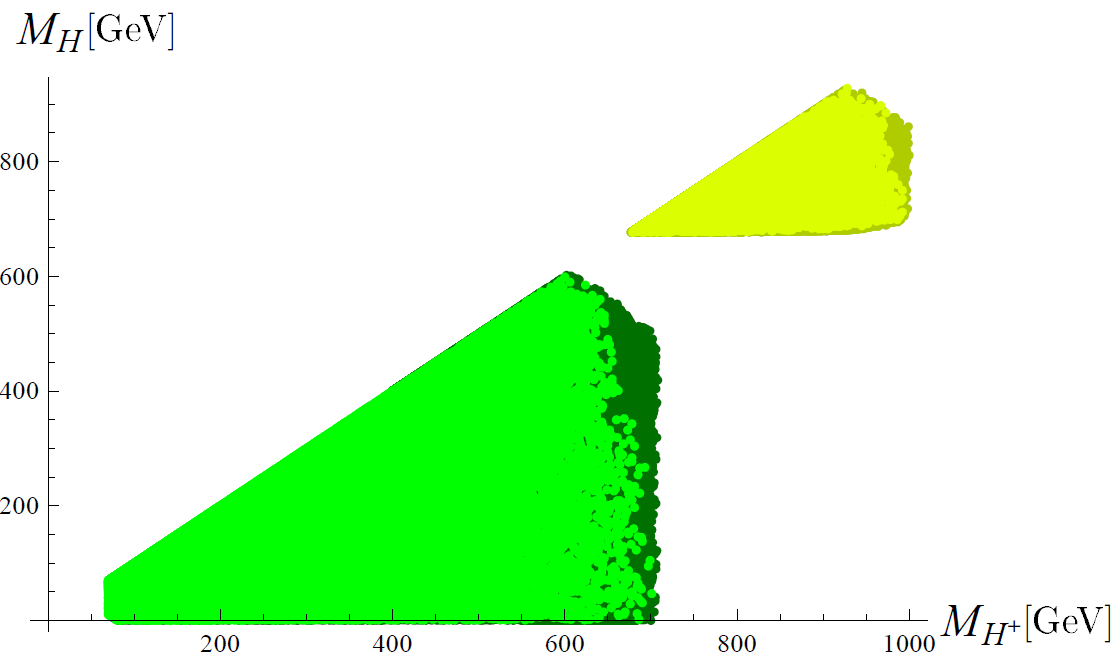}
\caption{Regions of masses allowed in IDM by  constraints (\ref{war}) [dark green (gray)] and  by~(\ref{war}),~(\ref{ewpt}) [light green (gray)]. In both of the plots the figures in the lower left corner correspond to the case with $m_{22}^2=0$, while the figures in the upper right corner (pale colors) correspond to $m_{22}^2=-10^6\g^2$.  \label{fig:inert}}
\end{figure}

It can be seen in Fig.~\ref{fig:inert} that if apart from the conditions~(\ref{war}) we also impose the EWPTs~(\ref{ewpt}), the pa\-ra\-me\-ter space is further constrained. Nevertheless, the overall bounds on the values of masses~(\ref{inertwyn}) remain unchanged.

As before, the bounds discussed above depend on $M_h$, and changing $M_h$ to 126~GeV would not visibly affect the result. Setting $M_h=200\g$ would shift the bounds on masses only slightly (up to 5\%). However for heavier $h$ smaller mass splittings between $H^{\pm}$ and $A$ would be allowed, only up to 100~GeV (to be compared with Fig.~\ref{fig:inert}, upper panel).


\subsection{Dark Matter mass}

IDM was shown~\cite{Ma:2007, Barbieri:2006, Dolle:2009, Gustafsson:2007,  *LopezHonorez:2006, *LopezHonorez:2007, *Tytgat:2007,  *Dolle-Miao:2009,  *Arina:2009, *Honorez:2010, *LopezHonorez:2010, *Sokolowska:2011, *Sokolowska:2011-acta} to accommodate a good DM candidate. In this work we assumed that $H$ is the light\-est $D$-odd particle and thus is the DM candidate. Three  regions of DM masses can be consistent with astrophysical data: $M_{\textrm{DM}}\lesssim 10\g$, $40\lesssim M_{\textrm{DM}}\lesssim 80\g$ or $M_{\textrm{DM}}\gtrsim 500\g$. Although the bounds on $M_H$ that we present do not affect DM considerations in the IDM in the low mass regimes, they do in the high mass region. If $H$ is in the high mass regime it can get only as heavy as $602\g$ [Eq.\,(\ref{inertwyn}) and Fig.\,\ref{fig:inert}]. Thus the high DM mass scenario is highly constrained. Of course this is the case only if we assume that $|m_{22}^2|$ is a reasonably small parameter. Otherwise, if $m_{22}^2\ll -10^4 \g^2$, $H$ can be very heavy, without violating perturbative unitarity. But then, all the dark scalars are very heavy as well (Fig.\,\ref{fig:inert}).


\subsection{New constraint on  $m_{22}^2$ parameter}

Although the parameter $m_{22}^2$ cannot be constrained directly by the unitarity condition, one can derive an upper bound on its value, taking into account the condition determining the existence of the Inert vacuum Eq.~(\ref{inert-war}). Expressing $m_{11}^2$ and $\lambda_1$ by the mass of the Higgs boson $m_{11}^2=M_h^2$, $\la=\frac{M_h^2}{v^2}$ and substituting to Eq.~(\ref{inert-war}) yields
$$m_{22}^2<\sqrt{\lb}M_h v.$$
Using the upper bound on $\lb$ [Eq.\,(\ref{wyniki lambdy})] and values of $M_h=125\g$ and $v=246\g$ we obtain the following limit on $m_{22}^2$:
$$
m_{22}^2\lesssim9\cdot 10^4\g^2.
$$
It is a new condition that has not been exploited before, since in many analyses the possibility of coexistence of different minima was overlooked.


\section{Mixed Model\label{sec:mixed}}

\subsection{Constraints on masses of the scalars}
Perturbative unitary condition for the Mixed Model has been analyzed in the past~\cite{Huffel:1981, Weldon:1984, Casalbuoni:1986, Casalbuoni:1988, Maalampi:1991, Kanemura:1993, Akeroyd:2000, Arhrib:2000, Ginzburg:2003, Ginzburg:2005, Horejsi:2005}, and both analytical and numerical bounds on scalar masses have been obtained. Our bounds following from conditions~(\ref{war}) read
\be\label{ogr-mix}
\begin{array}{r@{\;\leqslant\;}l}
M_H & 697\g,\\*
\m & 707\g,\\*
M_A & 706\g,\\*
M_h & 446\g.\\*
\end{array}
\ee
They are in good agreement with the most precise ana\-lyti\-cal results of Ref.~\cite{Horejsi:2005}.\footnote{Small discrepancies  are due to the uncertainty of the numerical method applied in our analysis.} Larger differences  appear between our results and results of Ref.~\cite{Akeroyd:2000, Arhrib:2000} (see also comparison in~\cite{Horejsi:2005}), which are probably due to uncertainties of the numerical method used in~\cite{Akeroyd:2000, Arhrib:2000}. 

On the constraints~(\ref{ogr-mix}), the bounds following from the EWPTs  can be superposed. The resulting regions of masses allowed by the conditions~(\ref{warunki}) are presented in Fig.~\ref{fig:mixed}. It is visible that EWPTs hardly change the upper bounds~(\ref{ogr-mix}).

\begin{figure}[ht]
\centering
\includegraphics[width=\columnwidth]{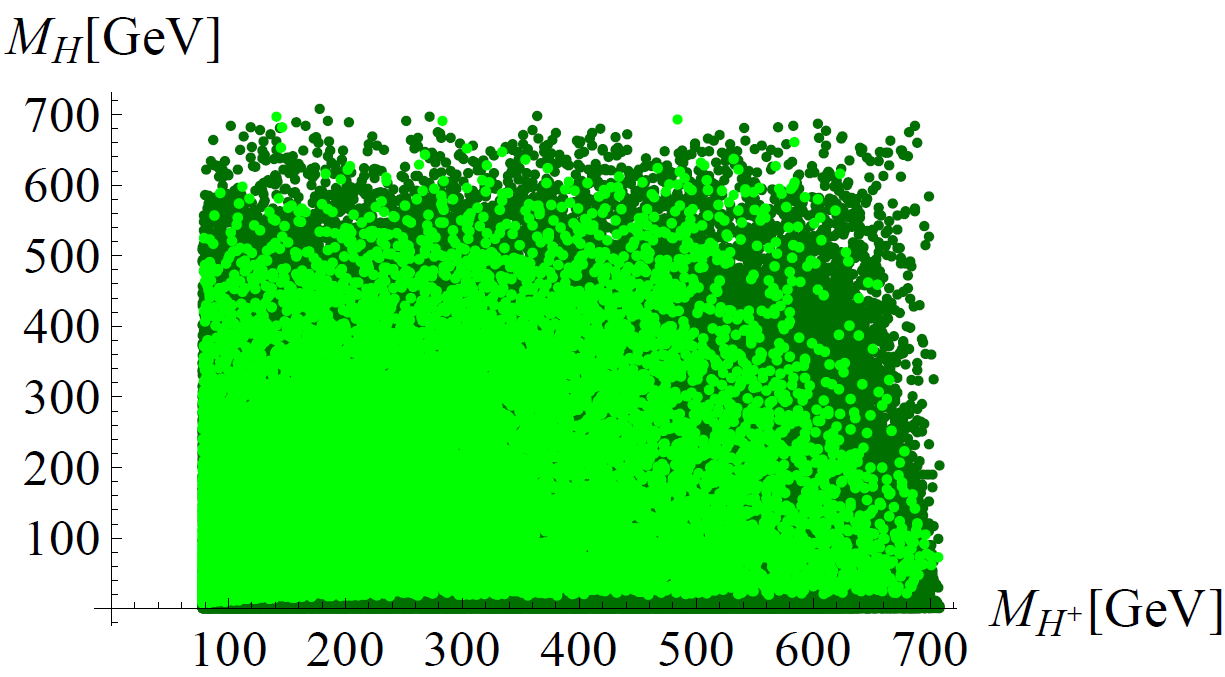}\\
\includegraphics[width=\columnwidth]{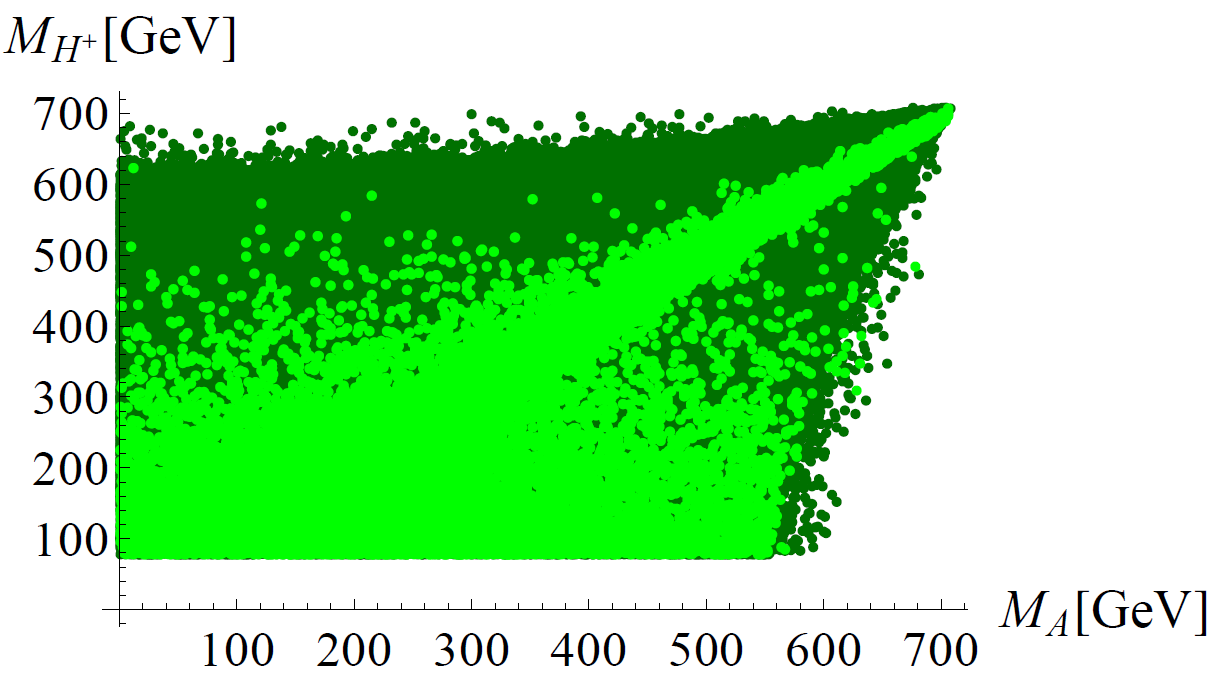}
\caption{Regions of masses allowed in the Mixed Model by the constraints (\ref{war}) (dark green/gray) and by~(\ref{war},~\ref{ewpt}) (light green/gray).\label{fig:mixed}}
\end{figure}



\subsection{New constraints on $\tan\beta$}

We also investigated  correlations between the values of scalars' masses and $\tan\beta$ allowed by the conditions~(\ref{war}), see also~\cite{Gorczyca:2011kr}.   $M_h$ exhibits interesting dependence on $\tan\beta$, the results are presented in Fig.\,\ref{fig:tbh}. It shows maximal allowed values of $M_h$ as a function of $\tan\beta$, $M_h^{\textrm{max}}(\tan\beta)$. We have checked that for any  value of $\sin(\beta-\alpha)$ only the area below that curve $M_h^{\textrm{max}}(\tan\beta)$ is allowed. From Fig.~\ref{fig:tbh} it follows that if we  consider a particular value of the mass of $h$  (or at least set a lower bound on it), then $\tan\beta$ is constrained, both from above and below. For example, for $M_h=125\g$  $0.5\lesssim\tan\beta\lesssim6.5$, $M_h=126\g$ would result in a similar bound. Shifting $M_h$ up would result in even stronger bounds. These bounds can be improved by fixing the mass of the Higgs boson explicitly in the code of the program and by imposing also the conditions following from EWPTs~(\ref{ewpt}). Then for $M_h=125\g$ the following bound is obtained:
\be\label{tb}
0.18\lesssim\tan\beta\lesssim5.59.
\ee
It should be underlined that this bound is obtained solely from the constraints~(\ref{warunki}) and the assumption $M_h=125\g$, without any assumptions on Yukawa couplings.\footnote{In particular, this bound is also true in 2HDM with Model I of Yukawa interactions.}

\begin{figure}[ht]
\centering
\includegraphics[width=\columnwidth]{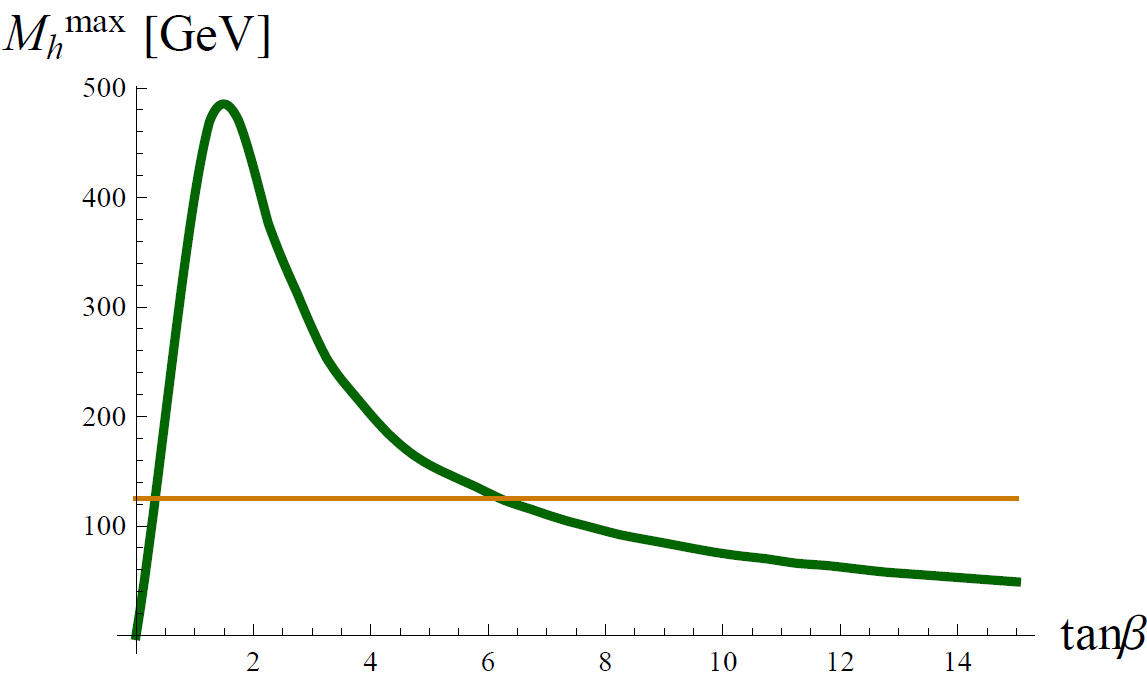}
\caption{Correlation between maximal values of the mass of $h$ and $\tan\beta$ in the Mixed Model allowed by constraints~(\ref{war}). For any value of $\sin(\beta-\alpha)$ only the points lying below the curve are allowed. The horizontal line corresponds to the mass of $125\g$.  \label{fig:tbh}}
\end{figure}

The correlations between $M_H$ and $\tan\beta$ are more complicated. For different values of $\sin(\beta-\alpha)$ we obtained different curves $M_H^{\textrm{max}}(\tan\beta)$. In particular for the case with $\sin(\beta-\alpha)=0$ the corresponding curve is just a straight line at $M_H\approx700\g$. This means that fixing $M_H$ cannot introduce any bounds on $\tan\beta$ unless we fix $\sin(\beta-\alpha)$ to a nonzero value. 

To understand why there is a bound on $\tan\beta$ in the SM-like scenario with $M_h=125\g$, while there is no bound in the $M_H=125\g$ case, one should go back to  the expressions for masses of the scalars, Eq.~(\ref{masymix}) \cite{Gorczyca:2011kr}. It can be easily checked that $M_h\to 0$ as $\tan\beta\to 0$ or $\tan\beta\to\infty$ for any fixed values of $\la$, $\lb$, $\lambda_{345}$.  Thus the curve in Fig.~\ref{fig:tbh} was bound to tend to zero for $\tan\beta\to 0$ and $\tan\beta\to\infty$, and so fixing $M_h$ must introduce an upper and a lower bound on $\tan\beta$. On the contrary, this does not apply to $M_H$ and therefore no bound on $\tan\beta$ follows from the assumption $M_H=125\g$ [independently of $\sin(\beta-\alpha)$].

\subsection{SM-like scenarios}

We can consider $h$ of the Mixed Model to be a candidate for the boson observed at the LHC. Then it is required to be SM-like, so $\sin(\beta-\alpha)$ has to be close to one. Of course the bound~(\ref{tb}) holds for such a case, so we conclude that in the SM-like 2HDM (Mixed), with $h$ playing the role of the SM Higgs boson with mass equal $125\g$, $\tan\beta=\frac{v_D}{v_S}$ is strongly constrained. Allowed regions of masses for this case [$M_h=125\g$, $\sin(\beta-\alpha)\geqslant0.98$] are presented in Fig.~\ref{SM-like-h}.
\begin{figure}[ht]
\centering
\includegraphics[width=\columnwidth]{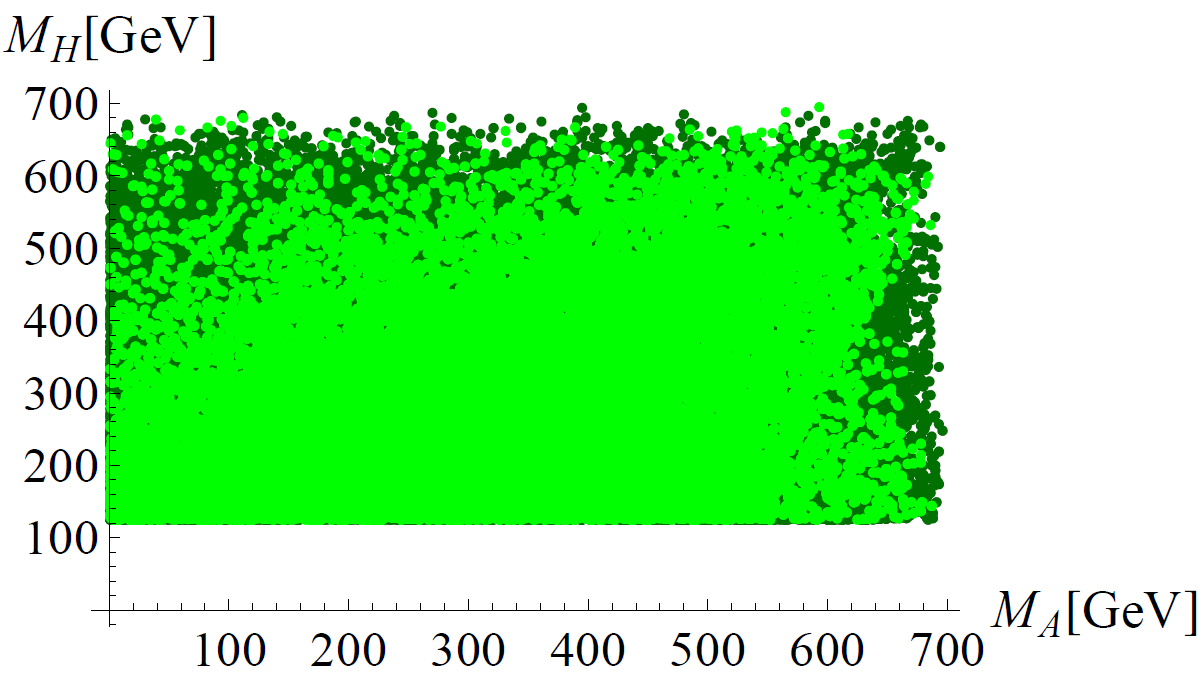}\\
\includegraphics[width=\columnwidth]{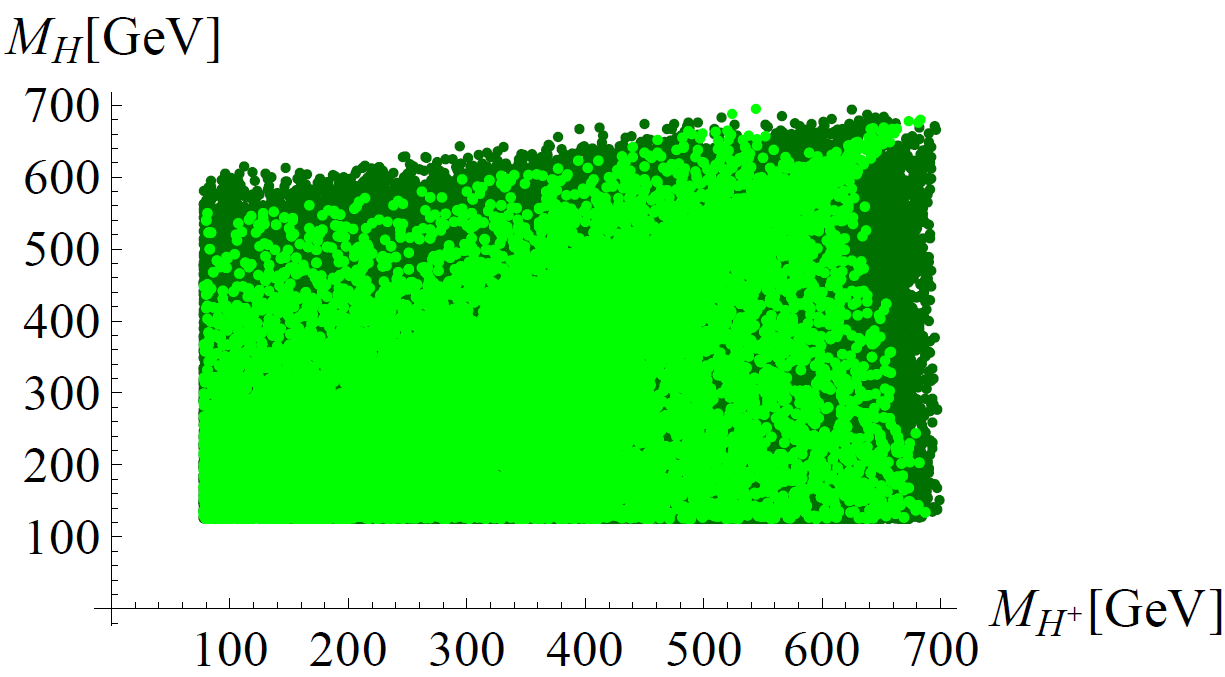}\\
\includegraphics[width=\columnwidth]{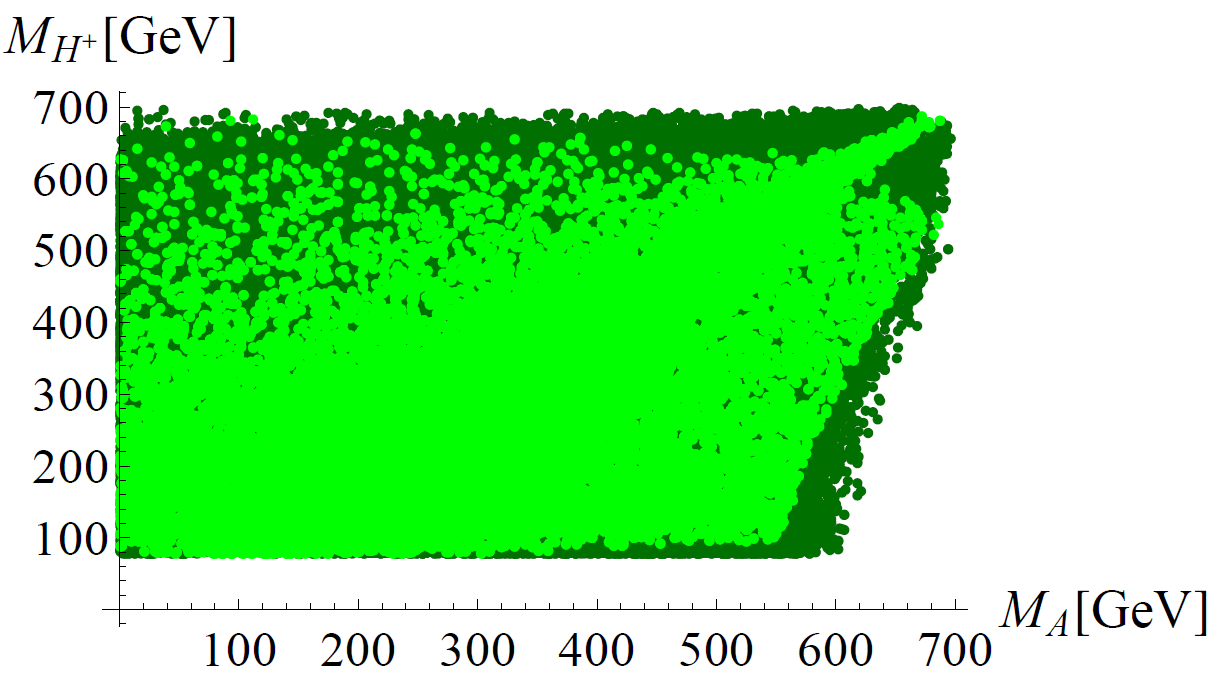}
\caption{Regions of masses allowed in the SM-like Mixed Model [with $M_h=125\g$ and $\sin(\beta-\alpha)\geqslant0.98$] by the constraints~(\ref{war}) [dark green (gray)] and by~(\ref{war}),~(\ref{ewpt}) [light green (gray)].  \label{SM-like-h}}
\end{figure}

We can also consider a SM-like  $H$  boson, with $\cos(\beta-\alpha)$  close to one [$\sin(\beta-\alpha)\approx0$]. Then, as a consequence of the results presented in the previous section, $\tan\beta$ is not constrained by conditions~(\ref{warunki}) and assumption on the mass of $H$. The regions of other scalars' masses allowed in this scenario [$M_H=125\g$ and $0\leqslant\sin(\beta-\alpha)\leqslant0.2$] are presented in Fig.~\ref{sm-like-Ha}. It can be observed that unless $H^{\pm}$ is light, with the mass up to $150$--$200\g$, $H^{\pm}$ and $A$ have to be approximately mass degenerate.
\begin{figure}[ht]
\centering
\includegraphics[width=\columnwidth]{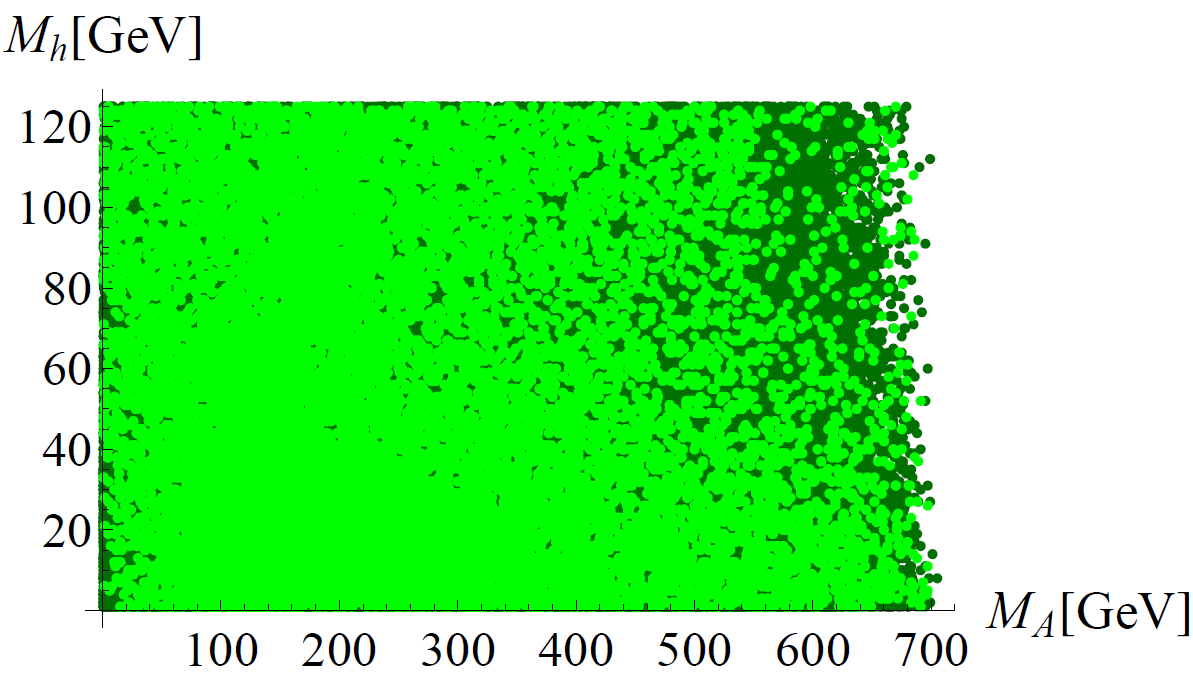}\\
\includegraphics[width=\columnwidth]{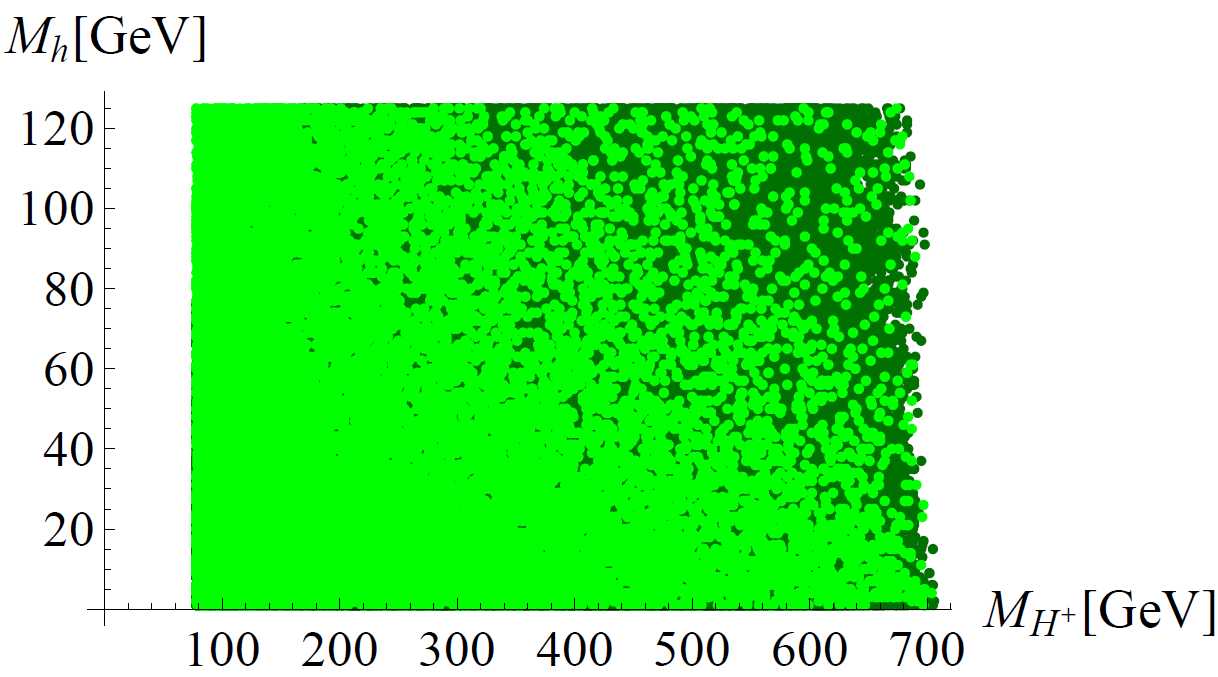}\\
\includegraphics[width=\columnwidth]{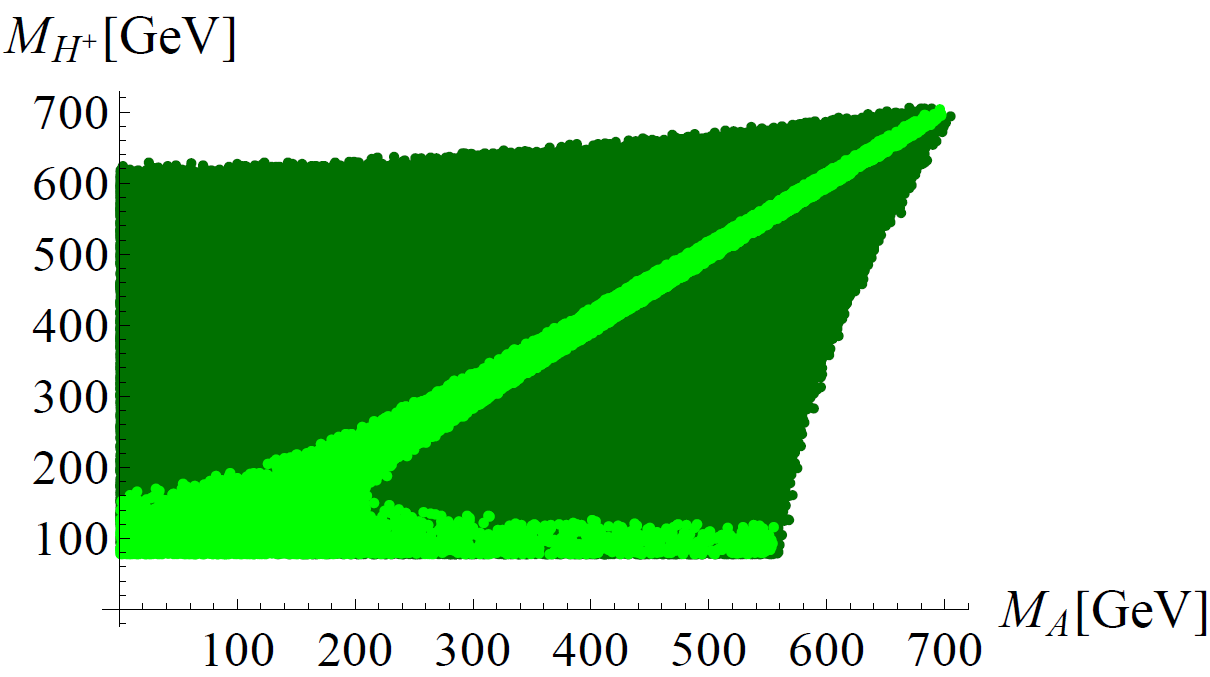}
\caption{Regions of masses allowed in the SM-like Mixed Model [with $M_H=125\g$ and $0\leqslant\sin(\beta-\alpha)\leqslant0.2$] by the constraints~(\ref{war}) [dark green (gray)] and by~(\ref{war}),~(\ref{ewpt}) [light green (gray)].  \label{sm-like-Ha}}
\end{figure}

The region in the $(M_h,\ M_A)$ plane could be further constrained using the LEP bounds~\cite{LEP-modelind}. Figure~6 in Ref.~\cite{LEP-modelind} shows excluded regions in the $(M_h,\ M_A)$ plane depending on the value of $\cos^2(\beta-\alpha)$. For $\sin(\beta-\alpha)>0.2$,  $\cos(\beta-\alpha)>0.96$ and with this value approximately the lower left corner, where $M_h<100\g-M_A$, would be excluded according to Ref.~\cite{LEP-modelind}.

\subsection{Discussion}

Although the constraint~(\ref{tb}) does not depend on the model of Yukawa interactions, it imposes limitations on them, as the couplings of fermions to the light Higgs boson are proportional to $\tan\beta$ or $\cot\beta$. For example, the constraint~(\ref{tb}) is far more stringent than the bound of the order of 100--200 coming from the assumption of perturbativity of the quark couplings presented in Ref.~\cite{Barger:1989} for Model II of Yukawa interactions. Furthermore, our constraint is important for the charged Higgs boson searches, because $\tan\beta$ governs the couplings between fermions and $H^{\pm}$. In Ref.~\cite{Guedes:2012}   it was shown that with the charged Higgs search data from LHC@14 TeV the part of the $(\tan\beta,\ \m)$ plane where $\tan\beta<6$ and $\m<150\g$ will be mostly excluded for the 2HDMs with Type~X and Type~I of Yukawa interactions. Combined with the bound~(\ref{tb}), this would rule out (or strongly disfavor) light charged Higgs in these models.

\section{Summary\label{sec:sum}}

We performed an analysis of the parameter spaces of two different kinds of  2HDMs with a $\mathbb{Z}_2$-symmetric potential. Using vacuum stability conditions, perturbative unitarity condition,  conditions determining type of the vacuum, the LEP bounds on scalars' masses as well as EWPTs we constrained the possible values of parameters. As a result we obtained the regions in the space of quartic parameters where the perturbative description is valid and consistent with theoretical assumptions. 

In the IDM we presented the regions in the spaces of quartic couplings and masses of the scalars which are allowed by the imposed conditions. Moreover we found a new type of upper limit on the mass parameter: $m_{22}^2\lesssim9\cdot 10^4\g^2$, following from the requirement that the Inert vacuum is a stable one.

For the Mixed Model we presented general viable regions in the space of scalars' masses. In the case with the Higgs boson $h$ of mass $125\g$ we found that $\tan\beta$ is strongly constrained, namely $0.18\lesssim\tan\beta\lesssim5.59$ and we presented allowed regions in the mass parameter space for the SM-like case [$\sin(\beta-\alpha)>0.98$].  We also examined the SM-like limit of the Mixed Model with $M_H=125\g$ and $0<\sin(\beta-\alpha)<0.2$ and found no bounds on $\tan\beta$. However for this case, either $H^{\pm}$ has to be fairly light or $H^{\pm}$ and $A$ have to have similar masses.

We stress the fact that the bound on $\tan\beta$ that we have found does not require any assumptions about the Yukawa sector. Nonetheless it constraints Yukawa interactions far more than the assumption of perturbativity of the Yukawa couplings and has important implications for the charged Higgs boson searches.

\begin{acknowledgments}
We are very grateful to M.~Krawczyk for advice, fruitful discussions and carefully reading the manuscript. We would like to thank D.~Sokołowska and D.~O'Neil for helpful discussions. The work was partly supported by Polish Ministry of Science and Higher Education Grant No. KBN N202 230337.
\end{acknowledgments}

\appendix

\section*{Formulas for $S$ and $T$ parameters\label{st-formulas}}
\subsubsection{Oblique parameters in the IDM}

For the IDM the expressions for $S$ and $T$ we have taken from \cite{Barbieri:2006} (see also \cite{Grimus:2008t, Arhrib:2012}\footnote{Note that in the formulas in~\cite{Grimus:2008, Arhrib:2012} terms containing $M_h$ and $M_{h_{\textrm{ref}}}$ do not appear. They are assumed to cancel, however it is not the case unless $M_h=M_{h_{\textrm{ref}}}$, which is not always true.}):

\begin{align}
T&=\frac{1}{32\pi^2\alpha v^2}\bigg(F(\m^2,M_A^2)+F(\m^2,M_H^2)\nonumber\\*
&\phantom{==}-F(M_A^2,M_H^2)\bigg)+\frac{3}{8\pi}\log\bigg(\frac{M_{h_{\textrm{ref}}}}{M_h}\bigg),\nonumber\\
S&=\frac{1}{2\pi}\Bigg\{\frac{1}{6}\frac{M_A^4\big(M_A^2-3M_H^2\big)}{(M_A^2-M_H^2)^3}\log\bigg(\frac{M_A^2}{M_H^2}\bigg)+\frac{1}{6}\log\bigg(\frac{M_H^2}{\m^2}\bigg)\nonumber\\
&\phantom{==}-\frac{5}{36} + \frac{M_H^2 M_A^2}{3\big(M_A^2-M_H^2\big)^2}-\frac{1}{3}\log\bigg(\frac{M_{h_{\textrm{ref}}}}{M_h}\bigg)\Bigg\}.\nonumber
\end{align}
$M_{h_{\textrm{ref}}}$ denotes the reference value of the SM Higgs boson mass used in the fit of the experimental data ($M_{h_{\textrm{ref}}}=117\g$).
Moreover, function $F$ is defined as follows:
$$
F(x,y)=
\begin{cases}
\frac{x+y}{2}-\frac{xy}{x-y}\log\frac{x}{y}& \textrm{for}\quad x\neq y,\\
0&\textrm{for}\quad x=y.\\
\end{cases}
$$

\subsubsection{Oblique parameters in the Mixed Model}

For the Mixed Model, using formulas from~\cite{Grimus:2008,  Grimus:2008t} adapted for a two-doublet case, the following expressions for the $S$ and $T$ parameters are obtained (compare also~\cite{Eriksson:2010}):
\begin{align}
T=&\frac{1}{16 \pi^2 v^2\alpha}\Big\{F(\m^2,M_A^2)\nonumber\\*
&+\sin^2(\beta-\alpha)\Big[F(\m^2,M_H^2)-F(M_A^2,M_H^2)\Big]\nonumber\\*
&+\cos^2(\beta-\alpha)\Big[F(\m^2,M_h^2)-F(M_A^2,M_h^2)\Big]\nonumber\\*
&+3\cos^2(\beta-\alpha)\Big[F(M_Z^2,M_H^2)-F(M_W^2,M_H^2)\Big]\nonumber\\*
&+3\sin^2(\beta-\alpha)\Big[F(M_Z^2,M_h^2)-F(M_W^2,M_h^2)\Big]\nonumber\\*
&-3\Big[F(M_Z^2,M_{h_{\textrm{ref}}}^2)-F(M_W^2,M_{h_{\textrm{ref}}}^2)\Big]\Big\}\nonumber,
\end{align}
\begin{align}
S=&\frac{1}{24\pi}\Big\{(2s_W^2 - 1)^2 G(\m^2,\m^2,M_Z^2)\nonumber\\*
& + \sin^2(\beta-\alpha)\Big[G(M_A^2,M_H^2,M_Z^2)+\hat{G}(M_h^2,M_Z^2)\Big]\nonumber\\*
&+\cos^2(\beta-\alpha)\Big[G(M_A^2,M_h^2,M_Z^2)+\hat{G}(M_H^2,M_Z^2)\Big]\nonumber\\*
&-2\log\m^2+\log M_A^2+\log M_H^2\nonumber\\*
&+ \log M_h^2 - \log M_{h_{\textrm{ref}}}^2 - \hat{G}(M_{h_{\textrm{ref}}},M_Z^2)\Big\}\nonumber.
\end{align}
The following definitions were used:
\begin{align}
G(x,&y,z)=-\frac{16}{3}+\frac{5(x+y)}{z}-\frac{2(x-y)^2}{z^2}\nonumber\\*
&+\frac{3}{z}\bigg(\frac{x^2+y^2}{x-y}-\frac{x^2-y^2}{z}+\frac{(x-y)^3}{3z^2}\bigg)\log\frac{x}{y}\nonumber\\* 
&+\frac{z^2-2z(x+y)+(x-y)^2}{z^3}\nonumber\\*
&\phantom{+}\times f\big(x+y-z,z^2-2z(x+y)+(x-y)^2\big),\nonumber
\end{align}
$$
f(x,y)=
\begin{cases}
\sqrt{y}\log\Big|\frac{x-\sqrt{y}}{x+\sqrt{y}}\Big|  & \textrm{for}\quad y>0,\\*
0   &  \textrm{for}\quad y=0,\\*
2\sqrt{-y}\arctan\frac{\sqrt{-y}}{x}  &\textrm{for}\quad y<0,\\*
\end{cases}
$$
\begin{align}
\hat{G}(x,y)=&-\frac{79}{3}+9\frac{x}{y}-2\frac{x^2}{y^2}\nonumber\\*
&+\bigg(-10+18\frac{x}{y}-6\frac{x^2}{y^2}+\frac{x^2}{y^3}-9\frac{x+y}{x-y}\bigg)\log\frac{x}{y}\nonumber\\*
&+\bigg(12-4\frac{x}{y}+\frac{x^2}{y^2}\bigg)\frac{f(x,x^2-4xy)}{y}.\nonumber
\end{align}

\vspace{\stretch{1}}

\bibliography{biblio}

\begin{thebibliography}{57}%
\makeatletter
\providecommand \@ifxundefined [1]{%
 \@ifx{#1\undefined}
}%
\providecommand \@ifnum [1]{%
 \ifnum #1\expandafter \@firstoftwo
 \else \expandafter \@secondoftwo
 \fi
}%
\providecommand \@ifx [1]{%
 \ifx #1\expandafter \@firstoftwo
 \else \expandafter \@secondoftwo
 \fi
}%
\providecommand \natexlab [1]{#1}%
\providecommand \enquote  [1]{``#1''}%
\providecommand \bibnamefont  [1]{#1}%
\providecommand \bibfnamefont [1]{#1}%
\providecommand \citenamefont [1]{#1}%
\providecommand \href@noop [0]{\@secondoftwo}%
\providecommand \href [0]{\begingroup \@sanitize@url \@href}%
\providecommand \@href[1]{\@@startlink{#1}\@@href}%
\providecommand \@@href[1]{\endgroup#1\@@endlink}%
\providecommand \@sanitize@url [0]{\catcode `\\12\catcode `\$12\catcode
  `\&12\catcode `\#12\catcode `\^12\catcode `\_12\catcode `\%12\relax}%
\providecommand \@@startlink[1]{}%
\providecommand \@@endlink[0]{}%
\providecommand \url  [0]{\begingroup\@sanitize@url \@url }%
\providecommand \@url [1]{\endgroup\@href {#1}{\urlprefix }}%
\providecommand \urlprefix  [0]{URL }%
\providecommand \Eprint [0]{\href }%
\providecommand \doibase [0]{http://dx.doi.org/}%
\providecommand \selectlanguage [0]{\@gobble}%
\providecommand \bibinfo  [0]{\@secondoftwo}%
\providecommand \bibfield  [0]{\@secondoftwo}%
\providecommand \translation [1]{[#1]}%
\providecommand \BibitemOpen [0]{}%
\providecommand \bibitemStop [0]{}%
\providecommand \bibitemNoStop [0]{.\EOS\space}%
\providecommand \EOS [0]{\spacefactor3000\relax}%
\providecommand \BibitemShut  [1]{\csname bibitem#1\endcsname}%
\let\auto@bib@innerbib\@empty
\bibitem [{\citenamefont {Branco}\ \emph {et~al.}(2012)\citenamefont {Branco},
  \citenamefont {Ferreira}, \citenamefont {Lavoura}, \citenamefont {Rebelo},
  \citenamefont {Sher},\ and\ \citenamefont {Silva}}]{Branco:2011}%
  \BibitemOpen
  \bibfield  {author} {\bibinfo {author} {\bibfnamefont {G.}~\bibnamefont
  {Branco}}, \bibinfo {author} {\bibfnamefont {P.}~\bibnamefont {Ferreira}},
  \bibinfo {author} {\bibfnamefont {L.}~\bibnamefont {Lavoura}}, \bibinfo
  {author} {\bibfnamefont {M.}~\bibnamefont {Rebelo}}, \bibinfo {author}
  {\bibfnamefont {M.}~\bibnamefont {Sher}}, \ and\ \bibinfo {author}
  {\bibfnamefont {J.}~\bibnamefont {Silva}},\ }\href {\doibase
  10.1016/j.physrep.2012.02.002} {\bibfield  {journal} {\bibinfo  {journal}
  {Phys.Rept.}\ }\textbf {\bibinfo {volume} {516}},\ \bibinfo {pages} {1}
  (\bibinfo {year} {2012})},\ \Eprint {http://arxiv.org/abs/1106.0034}
  {arXiv:1106.0034 [hep-ph]} \BibitemShut {NoStop}%
\bibitem [{\citenamefont {Deshpande}\ and\ \citenamefont {Ma}(1978)}]{Ma:1978}%
  \BibitemOpen
  \bibfield  {author} {\bibinfo {author} {\bibfnamefont {N.~G.}\ \bibnamefont
  {Deshpande}}\ and\ \bibinfo {author} {\bibfnamefont {E.}~\bibnamefont {Ma}},\
  }\href {\doibase 10.1103/PhysRevD.18.2574} {\bibfield  {journal} {\bibinfo
  {journal} {Phys.Rev.}\ }\textbf {\bibinfo {volume} {D18}},\ \bibinfo {pages}
  {2574} (\bibinfo {year} {1978})}\BibitemShut {NoStop}%
\bibitem [{\citenamefont {Barbieri}\ \emph {et~al.}(2006)\citenamefont
  {Barbieri}, \citenamefont {Hall},\ and\ \citenamefont
  {Rychkov}}]{Barbieri:2006}%
  \BibitemOpen
  \bibfield  {author} {\bibinfo {author} {\bibfnamefont {R.}~\bibnamefont
  {Barbieri}}, \bibinfo {author} {\bibfnamefont {L.~J.}\ \bibnamefont {Hall}},
  \ and\ \bibinfo {author} {\bibfnamefont {V.~S.}\ \bibnamefont {Rychkov}},\
  }\href {\doibase 10.1103/PhysRevD.74.015007} {\bibfield  {journal} {\bibinfo
  {journal} {Phys.Rev.}\ }\textbf {\bibinfo {volume} {D74}},\ \bibinfo {pages}
  {015007} (\bibinfo {year} {2006})},\ \Eprint
  {http://arxiv.org/abs/hep-ph/0603188} {arXiv:hep-ph/0603188 [hep-ph]}
  \BibitemShut {NoStop}%
\bibitem [{\citenamefont {Cao}\ \emph {et~al.}(2007)\citenamefont {Cao},
  \citenamefont {Ma},\ and\ \citenamefont {Rajasekaran}}]{Ma:2007}%
  \BibitemOpen
  \bibfield  {author} {\bibinfo {author} {\bibfnamefont {Q.-H.}\ \bibnamefont
  {Cao}}, \bibinfo {author} {\bibfnamefont {E.}~\bibnamefont {Ma}}, \ and\
  \bibinfo {author} {\bibfnamefont {G.}~\bibnamefont {Rajasekaran}},\ }\href
  {\doibase 10.1103/PhysRevD.76.095011} {\bibfield  {journal} {\bibinfo
  {journal} {Phys.Rev.}\ }\textbf {\bibinfo {volume} {D76}},\ \bibinfo {pages}
  {095011} (\bibinfo {year} {2007})},\ \Eprint {http://arxiv.org/abs/0708.2939}
  {arXiv:0708.2939 [hep-ph]} \BibitemShut {NoStop}%
\bibitem [{\citenamefont {Aad}\ \emph {et~al.}(2012)\citenamefont {Aad} \emph
  {et~al.}}]{atlas:2012}%
  \BibitemOpen
  \bibfield  {author} {\bibinfo {author} {\bibfnamefont {G.}~\bibnamefont
  {Aad}} \emph {et~al.} (\bibinfo {collaboration} {ATLAS Collaboration}),\
  }\href {\doibase 10.1016/j.physletb.2012.08.020} {\bibfield  {journal}
  {\bibinfo  {journal} {Phys.Lett.}\ }\textbf {\bibinfo {volume} {B716}},\
  \bibinfo {pages} {1} (\bibinfo {year} {2012})},\ \Eprint
  {http://arxiv.org/abs/1207.7214} {arXiv:1207.7214 [hep-ex]} \BibitemShut
  {NoStop}%
\bibitem [{\citenamefont {Chatrchyan}\ \emph {et~al.}(2012)\citenamefont
  {Chatrchyan} \emph {et~al.}}]{cms:2012}%
  \BibitemOpen
  \bibfield  {author} {\bibinfo {author} {\bibfnamefont {S.}~\bibnamefont
  {Chatrchyan}} \emph {et~al.} (\bibinfo {collaboration} {CMS Collaboration}),\
  }\href {\doibase 10.1016/j.physletb.2012.08.021} {\bibfield  {journal}
  {\bibinfo  {journal} {Phys.Lett.}\ }\textbf {\bibinfo {volume} {B716}},\
  \bibinfo {pages} {30} (\bibinfo {year} {2012})},\ \Eprint
  {http://arxiv.org/abs/1207.7235} {arXiv:1207.7235 [hep-ex]} \BibitemShut
  {NoStop}%
\bibitem [{\citenamefont {Dicus}\ and\ \citenamefont
  {Mathur}(1973)}]{Dicus:1973}%
  \BibitemOpen
  \bibfield  {author} {\bibinfo {author} {\bibfnamefont {D.~A.}\ \bibnamefont
  {Dicus}}\ and\ \bibinfo {author} {\bibfnamefont {V.~S.}\ \bibnamefont
  {Mathur}},\ }\href {\doibase 10.1103/PhysRevD.7.3111} {\bibfield  {journal}
  {\bibinfo  {journal} {Phys. Rev. D}\ }\textbf {\bibinfo {volume} {7}},\
  \bibinfo {pages} {3111} (\bibinfo {year} {1973})}\BibitemShut {NoStop}%
\bibitem [{\citenamefont {Lee}\ \emph {et~al.}(1977{\natexlab{a}})\citenamefont
  {Lee}, \citenamefont {Quigg},\ and\ \citenamefont {Thacker}}]{Lee:1977prl}%
  \BibitemOpen
  \bibfield  {author} {\bibinfo {author} {\bibfnamefont {B.~W.}\ \bibnamefont
  {Lee}}, \bibinfo {author} {\bibfnamefont {C.}~\bibnamefont {Quigg}}, \ and\
  \bibinfo {author} {\bibfnamefont {H.}~\bibnamefont {Thacker}},\ }\href
  {\doibase 10.1103/PhysRevLett.38.883} {\bibfield  {journal} {\bibinfo
  {journal} {Phys.Rev.Lett.}\ }\textbf {\bibinfo {volume} {38}},\ \bibinfo
  {pages} {883} (\bibinfo {year} {1977}{\natexlab{a}})}\BibitemShut {NoStop}%
\bibitem [{\citenamefont {Lee}\ \emph {et~al.}(1977{\natexlab{b}})\citenamefont
  {Lee}, \citenamefont {Quigg},\ and\ \citenamefont {Thacker}}]{Lee:1977}%
  \BibitemOpen
  \bibfield  {author} {\bibinfo {author} {\bibfnamefont {B.~W.}\ \bibnamefont
  {Lee}}, \bibinfo {author} {\bibfnamefont {C.}~\bibnamefont {Quigg}}, \ and\
  \bibinfo {author} {\bibfnamefont {H.}~\bibnamefont {Thacker}},\ }\href
  {\doibase 10.1103/PhysRevD.16.1519} {\bibfield  {journal} {\bibinfo
  {journal} {Phys.Rev.}\ }\textbf {\bibinfo {volume} {D16}},\ \bibinfo {pages}
  {1519} (\bibinfo {year} {1977}{\natexlab{b}})}\BibitemShut {NoStop}%
\bibitem [{\citenamefont {Weldon}(1984)}]{Weldon:1984}%
  \BibitemOpen
  \bibfield  {author} {\bibinfo {author} {\bibfnamefont {H.~A.}\ \bibnamefont
  {Weldon}},\ }\href {\doibase 10.1103/PhysRevD.30.1547} {\bibfield  {journal}
  {\bibinfo  {journal} {Phys.Rev.}\ }\textbf {\bibinfo {volume} {D30}},\
  \bibinfo {pages} {1547} (\bibinfo {year} {1984})}\BibitemShut {NoStop}%
\bibitem [{\citenamefont {Huffel}\ and\ \citenamefont
  {Pocsik}(1981)}]{Huffel:1981}%
  \BibitemOpen
  \bibfield  {author} {\bibinfo {author} {\bibfnamefont {H.}~\bibnamefont
  {Huffel}}\ and\ \bibinfo {author} {\bibfnamefont {G.}~\bibnamefont
  {Pocsik}},\ }\href {\doibase 10.1007/BF01429824} {\bibfield  {journal}
  {\bibinfo  {journal} {Z.Phys.}\ }\textbf {\bibinfo {volume} {C8}},\ \bibinfo
  {pages} {13} (\bibinfo {year} {1981})}\BibitemShut {NoStop}%
\bibitem [{\citenamefont {Casalbuoni}\ \emph {et~al.}(1986)\citenamefont
  {Casalbuoni}, \citenamefont {Dominici}, \citenamefont {Gatto},\ and\
  \citenamefont {Giunti}}]{Casalbuoni:1986}%
  \BibitemOpen
  \bibfield  {author} {\bibinfo {author} {\bibfnamefont {R.}~\bibnamefont
  {Casalbuoni}}, \bibinfo {author} {\bibfnamefont {D.}~\bibnamefont
  {Dominici}}, \bibinfo {author} {\bibfnamefont {R.}~\bibnamefont {Gatto}}, \
  and\ \bibinfo {author} {\bibfnamefont {C.}~\bibnamefont {Giunti}},\ }\href
  {\doibase 10.1016/0370-2693(86)91502-9} {\bibfield  {journal} {\bibinfo
  {journal} {Phys.Lett.}\ }\textbf {\bibinfo {volume} {B178}},\ \bibinfo
  {pages} {235} (\bibinfo {year} {1986})}\BibitemShut {NoStop}%
\bibitem [{\citenamefont {Casalbuoni}\ \emph {et~al.}(1988)\citenamefont
  {Casalbuoni}, \citenamefont {Dominici}, \citenamefont {Feruglio},\ and\
  \citenamefont {Gatto}}]{Casalbuoni:1988}%
  \BibitemOpen
  \bibfield  {author} {\bibinfo {author} {\bibfnamefont {R.}~\bibnamefont
  {Casalbuoni}}, \bibinfo {author} {\bibfnamefont {D.}~\bibnamefont
  {Dominici}}, \bibinfo {author} {\bibfnamefont {F.}~\bibnamefont {Feruglio}},
  \ and\ \bibinfo {author} {\bibfnamefont {R.}~\bibnamefont {Gatto}},\ }\href
  {\doibase 10.1016/0550-3213(88)90469-5} {\bibfield  {journal} {\bibinfo
  {journal} {Nucl.Phys.}\ }\textbf {\bibinfo {volume} {B299}},\ \bibinfo
  {pages} {117} (\bibinfo {year} {1988})}\BibitemShut {NoStop}%
\bibitem [{\citenamefont {Maalampi}\ \emph {et~al.}(1991)\citenamefont
  {Maalampi}, \citenamefont {Sirkka},\ and\ \citenamefont
  {Vilja}}]{Maalampi:1991}%
  \BibitemOpen
  \bibfield  {author} {\bibinfo {author} {\bibfnamefont {J.}~\bibnamefont
  {Maalampi}}, \bibinfo {author} {\bibfnamefont {J.}~\bibnamefont {Sirkka}}, \
  and\ \bibinfo {author} {\bibfnamefont {I.}~\bibnamefont {Vilja}},\ }\href
  {\doibase 10.1016/0370-2693(91)90068-2} {\bibfield  {journal} {\bibinfo
  {journal} {Phys.Lett.}\ }\textbf {\bibinfo {volume} {B265}},\ \bibinfo
  {pages} {371} (\bibinfo {year} {1991})}\BibitemShut {NoStop}%
\bibitem [{\citenamefont {Kanemura}\ \emph {et~al.}(1993)\citenamefont
  {Kanemura}, \citenamefont {Kubota},\ and\ \citenamefont
  {Takasugi}}]{Kanemura:1993}%
  \BibitemOpen
  \bibfield  {author} {\bibinfo {author} {\bibfnamefont {S.}~\bibnamefont
  {Kanemura}}, \bibinfo {author} {\bibfnamefont {T.}~\bibnamefont {Kubota}}, \
  and\ \bibinfo {author} {\bibfnamefont {E.}~\bibnamefont {Takasugi}},\ }\href
  {\doibase 10.1016/0370-2693(93)91205-2} {\bibfield  {journal} {\bibinfo
  {journal} {Phys.Lett.}\ }\textbf {\bibinfo {volume} {B313}},\ \bibinfo
  {pages} {155} (\bibinfo {year} {1993})},\ \Eprint
  {http://arxiv.org/abs/hep-ph/9303263} {arXiv:hep-ph/9303263 [hep-ph]}
  \BibitemShut {NoStop}%
\bibitem [{\citenamefont {Akeroyd}\ \emph {et~al.}(2000)\citenamefont
  {Akeroyd}, \citenamefont {Arhrib},\ and\ \citenamefont
  {Naimi}}]{Akeroyd:2000}%
  \BibitemOpen
  \bibfield  {author} {\bibinfo {author} {\bibfnamefont {A.~G.}\ \bibnamefont
  {Akeroyd}}, \bibinfo {author} {\bibfnamefont {A.}~\bibnamefont {Arhrib}}, \
  and\ \bibinfo {author} {\bibfnamefont {E.-M.}\ \bibnamefont {Naimi}},\ }\href
  {\doibase 10.1016/S0370-2693(00)00962-X} {\bibfield  {journal} {\bibinfo
  {journal} {Phys.Lett.}\ }\textbf {\bibinfo {volume} {B490}},\ \bibinfo
  {pages} {119} (\bibinfo {year} {2000})},\ \Eprint
  {http://arxiv.org/abs/hep-ph/0006035} {arXiv:hep-ph/0006035 [hep-ph]}
  \BibitemShut {NoStop}%
\bibitem [{\citenamefont {Arhrib}(2000)}]{Arhrib:2000}%
  \BibitemOpen
  \bibfield  {author} {\bibinfo {author} {\bibfnamefont {A.}~\bibnamefont
  {Arhrib}},\ }\href@noop {} {\  (\bibinfo {year} {2000})},\ \Eprint
  {http://arxiv.org/abs/hep-ph/0012353} {arXiv:hep-ph/0012353 [hep-ph]}
  \BibitemShut {NoStop}%
\bibitem [{\citenamefont {Ginzburg}\ and\ \citenamefont
  {Ivanov}(2003)}]{Ginzburg:2003}%
  \BibitemOpen
  \bibfield  {author} {\bibinfo {author} {\bibfnamefont {I.}~\bibnamefont
  {Ginzburg}}\ and\ \bibinfo {author} {\bibfnamefont {I.}~\bibnamefont
  {Ivanov}},\ }\href@noop {} {\  (\bibinfo {year} {2003})},\ \Eprint
  {http://arxiv.org/abs/hep-ph/0312374} {arXiv:hep-ph/0312374 [hep-ph]}
  \BibitemShut {NoStop}%
\bibitem [{\citenamefont {Ginzburg}\ and\ \citenamefont
  {Ivanov}(2005)}]{Ginzburg:2005}%
  \BibitemOpen
  \bibfield  {author} {\bibinfo {author} {\bibfnamefont {I.}~\bibnamefont
  {Ginzburg}}\ and\ \bibinfo {author} {\bibfnamefont {I.}~\bibnamefont
  {Ivanov}},\ }\href {\doibase 10.1103/PhysRevD.72.115010} {\bibfield
  {journal} {\bibinfo  {journal} {Phys.Rev.}\ }\textbf {\bibinfo {volume}
  {D72}},\ \bibinfo {pages} {115010} (\bibinfo {year} {2005})},\ \Eprint
  {http://arxiv.org/abs/hep-ph/0508020} {arXiv:hep-ph/0508020 [hep-ph]}
  \BibitemShut {NoStop}%
\bibitem [{\citenamefont {Ho\v{r}ej\v{s}\'i}\ and\ \citenamefont
  {Kladiva}(2006)}]{Horejsi:2005}%
  \BibitemOpen
  \bibfield  {author} {\bibinfo {author} {\bibfnamefont {J.}~\bibnamefont
  {Ho\v{r}ej\v{s}\'i}}\ and\ \bibinfo {author} {\bibfnamefont {M.}~\bibnamefont
  {Kladiva}},\ }\href {\doibase 10.1140/epjc/s2006-02472-3} {\bibfield
  {journal} {\bibinfo  {journal} {Eur.Phys.J.}\ }\textbf {\bibinfo {volume}
  {C46}},\ \bibinfo {pages} {81} (\bibinfo {year} {2006})},\ \Eprint
  {http://arxiv.org/abs/hep-ph/0510154} {arXiv:hep-ph/0510154 [hep-ph]}
  \BibitemShut {NoStop}%
\bibitem [{\citenamefont {Kanemura}\ \emph {et~al.}(2011)\citenamefont
  {Kanemura}, \citenamefont {Okada}, \citenamefont {Taniguchi},\ and\
  \citenamefont {Tsumura}}]{Kanemura:2011}%
  \BibitemOpen
  \bibfield  {author} {\bibinfo {author} {\bibfnamefont {S.}~\bibnamefont
  {Kanemura}}, \bibinfo {author} {\bibfnamefont {Y.}~\bibnamefont {Okada}},
  \bibinfo {author} {\bibfnamefont {H.}~\bibnamefont {Taniguchi}}, \ and\
  \bibinfo {author} {\bibfnamefont {K.}~\bibnamefont {Tsumura}},\ }\href
  {\doibase 10.1016/j.physletb.2011.09.035} {\bibfield  {journal} {\bibinfo
  {journal} {Phys.Lett.}\ }\textbf {\bibinfo {volume} {B704}},\ \bibinfo
  {pages} {303} (\bibinfo {year} {2011})},\ \Eprint
  {http://arxiv.org/abs/1108.3297} {arXiv:1108.3297 [hep-ph]} \BibitemShut
  {NoStop}%
\bibitem [{\citenamefont {Cheon}\ and\ \citenamefont {Kang}(2013)}]{Kang:2012}%
  \BibitemOpen
  \bibfield  {author} {\bibinfo {author} {\bibfnamefont {H.}~\bibnamefont
  {Cheon}}\ and\ \bibinfo {author} {\bibfnamefont {S.~K.}\ \bibnamefont
  {Kang}},\ }\href {\doibase 10.1007/JHEP09(2013)085} {\bibfield  {journal}
  {\bibinfo  {journal} {JHEP}\ }\textbf {\bibinfo {volume} {1309}},\ \bibinfo
  {pages} {085} (\bibinfo {year} {2013})},\ \Eprint
  {http://arxiv.org/abs/1207.1083} {arXiv:1207.1083 [hep-ph]} \BibitemShut
  {NoStop}%
\bibitem [{\citenamefont {Funk}\ \emph {et~al.}(2012)\citenamefont {Funk},
  \citenamefont {O'Neil},\ and\ \citenamefont {Winters}}]{Oneil:2012}%
  \BibitemOpen
  \bibfield  {author} {\bibinfo {author} {\bibfnamefont {G.}~\bibnamefont
  {Funk}}, \bibinfo {author} {\bibfnamefont {D.}~\bibnamefont {O'Neil}}, \ and\
  \bibinfo {author} {\bibfnamefont {R.~M.}\ \bibnamefont {Winters}},\ }\href
  {\doibase 10.1142/S0217751X12500212} {\bibfield  {journal} {\bibinfo
  {journal} {Int.J.Mod.Phys.}\ }\textbf {\bibinfo {volume} {A27}},\ \bibinfo
  {pages} {1250021} (\bibinfo {year} {2012})},\ \Eprint
  {http://arxiv.org/abs/1110.3812} {arXiv:1110.3812 [hep-ph]} \BibitemShut
  {NoStop}%
\bibitem [{\citenamefont {Gorczyca}(2011)}]{praca-mag}%
  \BibitemOpen
  \bibfield  {author} {\bibinfo {author} {\bibfnamefont {B.}~\bibnamefont
  {Gorczyca}},\ }\href@noop {} {\enquote {\bibinfo {title} {{Unitarity
  constraints for the Inert Doublet Model (in Polish)}},}\ }\bibinfo
  {howpublished} {{Master Thesis at the University of Warsaw}} (\bibinfo {year}
  {2011})\BibitemShut {NoStop}%
\bibitem [{\citenamefont {Gorczyca}\ and\ \citenamefont
  {Krawczyk}(2011{\natexlab{a}})}]{Gorczyca:2011}%
  \BibitemOpen
  \bibfield  {author} {\bibinfo {author} {\bibfnamefont {B.}~\bibnamefont
  {Gorczyca}}\ and\ \bibinfo {author} {\bibfnamefont {M.}~\bibnamefont
  {Krawczyk}},\ }\href {\doibase 10.5506/APhysPolB.43.481,
  10.5506/APhysPolB.42.2229} {\bibfield  {journal} {\bibinfo  {journal} {Acta
  Phys.Polon.}\ }\textbf {\bibinfo {volume} {B42}},\ \bibinfo {pages} {2229}
  (\bibinfo {year} {2011}{\natexlab{a}})},\ \Eprint
  {http://arxiv.org/abs/1112.4356} {arXiv:1112.4356 [hep-ph]} \BibitemShut
  {NoStop}%
\bibitem [{\citenamefont {Arhrib}\ \emph {et~al.}(2012)\citenamefont {Arhrib},
  \citenamefont {Benbrik},\ and\ \citenamefont {Gaur}}]{Arhrib:2012}%
  \BibitemOpen
  \bibfield  {author} {\bibinfo {author} {\bibfnamefont {A.}~\bibnamefont
  {Arhrib}}, \bibinfo {author} {\bibfnamefont {R.}~\bibnamefont {Benbrik}}, \
  and\ \bibinfo {author} {\bibfnamefont {N.}~\bibnamefont {Gaur}},\ }\href
  {\doibase 10.1103/PhysRevD.85.095021} {\bibfield  {journal} {\bibinfo
  {journal} {Phys.Rev.}\ }\textbf {\bibinfo {volume} {D85}},\ \bibinfo {pages}
  {095021} (\bibinfo {year} {2012})},\ \Eprint {http://arxiv.org/abs/1201.2644}
  {arXiv:1201.2644 [hep-ph]} \BibitemShut {NoStop}%
\bibitem [{\citenamefont {Gustafsson}\ \emph {et~al.}(2012)\citenamefont
  {Gustafsson}, \citenamefont {Rydbeck}, \citenamefont {Lopez-Honorez},\ and\
  \citenamefont {Lundstr\"om}}]{Gustafsson:2012}%
  \BibitemOpen
  \bibfield  {author} {\bibinfo {author} {\bibfnamefont {M.}~\bibnamefont
  {Gustafsson}}, \bibinfo {author} {\bibfnamefont {S.}~\bibnamefont {Rydbeck}},
  \bibinfo {author} {\bibfnamefont {L.}~\bibnamefont {Lopez-Honorez}}, \ and\
  \bibinfo {author} {\bibfnamefont {E.}~\bibnamefont {Lundstr\"om}},\ }\href
  {\doibase 10.1103/PhysRevD.86.075019} {\bibfield  {journal} {\bibinfo
  {journal} {Phys. Rev. D}\ }\textbf {\bibinfo {volume} {86}},\ \bibinfo
  {pages} {075019} (\bibinfo {year} {2012})}\BibitemShut {NoStop}%
\bibitem [{\citenamefont {Dolle}\ and\ \citenamefont {Su}(2009)}]{Dolle:2009}%
  \BibitemOpen
  \bibfield  {author} {\bibinfo {author} {\bibfnamefont {E.~M.}\ \bibnamefont
  {Dolle}}\ and\ \bibinfo {author} {\bibfnamefont {S.}~\bibnamefont {Su}},\
  }\href {\doibase 10.1103/PhysRevD.80.055012} {\bibfield  {journal} {\bibinfo
  {journal} {Phys.Rev.}\ }\textbf {\bibinfo {volume} {D80}},\ \bibinfo {pages}
  {055012} (\bibinfo {year} {2009})},\ \Eprint {http://arxiv.org/abs/0906.1609}
  {arXiv:0906.1609 [hep-ph]} \BibitemShut {NoStop}%
\bibitem [{\citenamefont {Ginzburg}\ \emph {et~al.}(2010)\citenamefont
  {Ginzburg}, \citenamefont {Kanishev}, \citenamefont {Krawczyk},\ and\
  \citenamefont {Sokolowska}}]{Krawczyk:2010}%
  \BibitemOpen
  \bibfield  {author} {\bibinfo {author} {\bibfnamefont {I.}~\bibnamefont
  {Ginzburg}}, \bibinfo {author} {\bibfnamefont {K.}~\bibnamefont {Kanishev}},
  \bibinfo {author} {\bibfnamefont {M.}~\bibnamefont {Krawczyk}}, \ and\
  \bibinfo {author} {\bibfnamefont {D.}~\bibnamefont {Sokolowska}},\ }\href
  {\doibase 10.1103/PhysRevD.82.123533} {\bibfield  {journal} {\bibinfo
  {journal} {Phys.Rev.}\ }\textbf {\bibinfo {volume} {D82}},\ \bibinfo {pages}
  {123533} (\bibinfo {year} {2010})},\ \Eprint {http://arxiv.org/abs/1009.4593}
  {arXiv:1009.4593 [hep-ph]} \BibitemShut {NoStop}%
\bibitem [{\citenamefont {Ginzburg}\ and\ \citenamefont
  {Krawczyk}(2005)}]{Krawczyk:2004sym}%
  \BibitemOpen
  \bibfield  {author} {\bibinfo {author} {\bibfnamefont {I.~F.}\ \bibnamefont
  {Ginzburg}}\ and\ \bibinfo {author} {\bibfnamefont {M.}~\bibnamefont
  {Krawczyk}},\ }\href {\doibase 10.1103/PhysRevD.72.115013} {\bibfield
  {journal} {\bibinfo  {journal} {Phys.Rev.}\ }\textbf {\bibinfo {volume}
  {D72}},\ \bibinfo {pages} {115013} (\bibinfo {year} {2005})},\ \Eprint
  {http://arxiv.org/abs/hep-ph/0408011} {arXiv:hep-ph/0408011 [hep-ph]}
  \BibitemShut {NoStop}%
\bibitem [{\citenamefont {Ginzburg}\ and\ \citenamefont
  {Krawczyk}()}]{Krawczyk:2004sym2}%
  \BibitemOpen
  \bibfield  {author} {\bibinfo {author} {\bibfnamefont {I.~F.}\ \bibnamefont
  {Ginzburg}}\ and\ \bibinfo {author} {\bibfnamefont {M.}~\bibnamefont
  {Krawczyk}},\ }\href@noop {} {}\bibinfo {howpublished} {{Proceedings of the
  XVIII International Workshop QFTHEP'2004}}\BibitemShut {NoStop}%
\bibitem [{\citenamefont {Branco}\ \emph {et~al.}(1999)\citenamefont {Branco},
  \citenamefont {Lavoura},\ and\ \citenamefont {Silva}}]{Branco:1999}%
  \BibitemOpen
  \bibfield  {author} {\bibinfo {author} {\bibfnamefont {G.~C.}\ \bibnamefont
  {Branco}}, \bibinfo {author} {\bibfnamefont {L.}~\bibnamefont {Lavoura}}, \
  and\ \bibinfo {author} {\bibfnamefont {J.~P.}\ \bibnamefont {Silva}},\
  }\href@noop {} {\emph {\bibinfo {title} {{CP Violation}}}}\ (\bibinfo
  {publisher} {Oxford University Press},\ \bibinfo {year} {1999})\BibitemShut
  {NoStop}%
\bibitem [{\citenamefont {Lundstrom}\ \emph {et~al.}(2009)\citenamefont
  {Lundstrom}, \citenamefont {Gustafsson},\ and\ \citenamefont
  {Edsjo}}]{Gustafsson:2009}%
  \BibitemOpen
  \bibfield  {author} {\bibinfo {author} {\bibfnamefont {E.}~\bibnamefont
  {Lundstrom}}, \bibinfo {author} {\bibfnamefont {M.}~\bibnamefont
  {Gustafsson}}, \ and\ \bibinfo {author} {\bibfnamefont {J.}~\bibnamefont
  {Edsjo}},\ }\href {\doibase 10.1103/PhysRevD.79.035013} {\bibfield  {journal}
  {\bibinfo  {journal} {Phys.Rev.}\ }\textbf {\bibinfo {volume} {D79}},\
  \bibinfo {pages} {035013} (\bibinfo {year} {2009})},\ \Eprint
  {http://arxiv.org/abs/0810.3924} {arXiv:0810.3924 [hep-ph]} \BibitemShut
  {NoStop}%
\bibitem [{\citenamefont {Gustafsson}(2010)}]{Gustafsson:2010}%
  \BibitemOpen
  \bibfield  {author} {\bibinfo {author} {\bibfnamefont {M.}~\bibnamefont
  {Gustafsson}},\ }\href@noop {} {\bibfield  {journal} {\bibinfo  {journal}
  {PoS}\ }\textbf {\bibinfo {volume} {CHARGED2010}},\ \bibinfo {pages} {030}
  (\bibinfo {year} {2010})},\ \Eprint {http://arxiv.org/abs/1106.1719}
  {arXiv:1106.1719 [hep-ph]} \BibitemShut {NoStop}%
\bibitem [{\citenamefont {Pierce}\ and\ \citenamefont
  {Thaler}(2007)}]{Pierce:2007}%
  \BibitemOpen
  \bibfield  {author} {\bibinfo {author} {\bibfnamefont {A.}~\bibnamefont
  {Pierce}}\ and\ \bibinfo {author} {\bibfnamefont {J.}~\bibnamefont
  {Thaler}},\ }\href {\doibase 10.1088/1126-6708/2007/08/026} {\bibfield
  {journal} {\bibinfo  {journal} {JHEP}\ }\textbf {\bibinfo {volume} {0708}},\
  \bibinfo {pages} {026} (\bibinfo {year} {2007})},\ \Eprint
  {http://arxiv.org/abs/hep-ph/0703056} {arXiv:hep-ph/0703056 [HEP-PH]}
  \BibitemShut {NoStop}%
\bibitem [{\citenamefont {Abbiendi}\ \emph {et~al.}(2013)\citenamefont
  {Abbiendi} \emph {et~al.}}]{Abbiendi:2013}%
  \BibitemOpen
  \bibfield  {author} {\bibinfo {author} {\bibfnamefont {G.}~\bibnamefont
  {Abbiendi}} \emph {et~al.} (\bibinfo {collaboration} {ALEPH, DELPHI, L3,
  OPAL, LEP}),\ }\href {\doibase 10.1140/epjc/s10052-013-2463-1} {\bibfield
  {journal} {\bibinfo  {journal} {Eur.Phys.J.}\ }\textbf {\bibinfo {volume}
  {C73}},\ \bibinfo {pages} {2463} (\bibinfo {year} {2013})},\ \Eprint
  {http://arxiv.org/abs/1301.6065} {arXiv:1301.6065 [hep-ex]} \BibitemShut
  {NoStop}%
\bibitem [{\citenamefont {Cornwall}\ \emph {et~al.}(1974)\citenamefont
  {Cornwall}, \citenamefont {Levin},\ and\ \citenamefont
  {Tiktopoulos}}]{Cornwall:1974}%
  \BibitemOpen
  \bibfield  {author} {\bibinfo {author} {\bibfnamefont {J.~M.}\ \bibnamefont
  {Cornwall}}, \bibinfo {author} {\bibfnamefont {D.~N.}\ \bibnamefont {Levin}},
  \ and\ \bibinfo {author} {\bibfnamefont {G.}~\bibnamefont {Tiktopoulos}},\
  }\href {\doibase 10.1103/PhysRevD.10.1145, 10.1103/PhysRevD.11.972}
  {\bibfield  {journal} {\bibinfo  {journal} {Phys.Rev.}\ }\textbf {\bibinfo
  {volume} {D10}},\ \bibinfo {pages} {1145} (\bibinfo {year}
  {1974})}\BibitemShut {NoStop}%
\bibitem [{\citenamefont {Grimus}\ \emph
  {et~al.}(2008{\natexlab{a}})\citenamefont {Grimus}, \citenamefont {Lavoura},
  \citenamefont {Ogreid},\ and\ \citenamefont {Osland}}]{Grimus:2008}%
  \BibitemOpen
  \bibfield  {author} {\bibinfo {author} {\bibfnamefont {W.}~\bibnamefont
  {Grimus}}, \bibinfo {author} {\bibfnamefont {L.}~\bibnamefont {Lavoura}},
  \bibinfo {author} {\bibfnamefont {O.}~\bibnamefont {Ogreid}}, \ and\ \bibinfo
  {author} {\bibfnamefont {P.}~\bibnamefont {Osland}},\ }\href {\doibase
  10.1016/j.nuclphysb.2008.04.019} {\bibfield  {journal} {\bibinfo  {journal}
  {Nucl.Phys.}\ }\textbf {\bibinfo {volume} {B801}},\ \bibinfo {pages} {81}
  (\bibinfo {year} {2008}{\natexlab{a}})},\ \Eprint
  {http://arxiv.org/abs/0802.4353} {arXiv:0802.4353 [hep-ph]} \BibitemShut
  {NoStop}%
\bibitem [{\citenamefont {Grimus}\ \emph
  {et~al.}(2008{\natexlab{b}})\citenamefont {Grimus}, \citenamefont {Lavoura},
  \citenamefont {Ogreid},\ and\ \citenamefont {Osland}}]{Grimus:2008t}%
  \BibitemOpen
  \bibfield  {author} {\bibinfo {author} {\bibfnamefont {W.}~\bibnamefont
  {Grimus}}, \bibinfo {author} {\bibfnamefont {L.}~\bibnamefont {Lavoura}},
  \bibinfo {author} {\bibfnamefont {O.}~\bibnamefont {Ogreid}}, \ and\ \bibinfo
  {author} {\bibfnamefont {P.}~\bibnamefont {Osland}},\ }\href {\doibase
  10.1088/0954-3899/35/7/075001} {\bibfield  {journal} {\bibinfo  {journal}
  {J.Phys.G}\ }\textbf {\bibinfo {volume} {G35}},\ \bibinfo {pages} {075001}
  (\bibinfo {year} {2008}{\natexlab{b}})},\ \Eprint
  {http://arxiv.org/abs/0711.4022} {arXiv:0711.4022 [hep-ph]} \BibitemShut
  {NoStop}%
\bibitem [{\citenamefont {Nakamura}\ \emph {et~al.}(2010)\citenamefont
  {Nakamura} \emph {et~al.}}]{Nakamura:2010}%
  \BibitemOpen
  \bibfield  {author} {\bibinfo {author} {\bibfnamefont {K.}~\bibnamefont
  {Nakamura}} \emph {et~al.} (\bibinfo {collaboration} {Particle Data Group}),\
  }\href {\doibase 10.1088/0954-3899/37/7A/075021} {\bibfield  {journal}
  {\bibinfo  {journal} {J.Phys.G}\ }\textbf {\bibinfo {volume} {G37}},\
  \bibinfo {pages} {075021} (\bibinfo {year} {2010})}\BibitemShut {NoStop}%
\bibitem [{\citenamefont {Beringer}\ \emph {et~al.}(2012)\citenamefont
  {Beringer} \emph {et~al.}}]{Beringer:2012}%
  \BibitemOpen
  \bibfield  {author} {\bibinfo {author} {\bibfnamefont {J.}~\bibnamefont
  {Beringer}} \emph {et~al.} (\bibinfo {collaboration} {Particle Data Group}),\
  }\href {\doibase 10.1103/PhysRevD.86.010001} {\bibfield  {journal} {\bibinfo
  {journal} {Phys. Rev. D}\ }\textbf {\bibinfo {volume} {86}},\ \bibinfo
  {pages} {010001} (\bibinfo {year} {2012})}\BibitemShut {NoStop}%
\bibitem [{\citenamefont {Winters}\ and\ \citenamefont {O'Neil}()}]{STelipse}%
  \BibitemOpen
  \bibfield  {author} {\bibinfo {author} {\bibfnamefont {M.~R.}\ \bibnamefont
  {Winters}}\ and\ \bibinfo {author} {\bibfnamefont {D.}~\bibnamefont
  {O'Neil}},\ }\href@noop {} {\enquote {\bibinfo {title}
  {{STellipse-Mathematica Implementation}},}\ }\bibinfo {howpublished}
  {\url{http://people.bridgewater.edu/~doneil/}}\BibitemShut {NoStop}%
\bibitem [{\citenamefont {Barger}\ \emph {et~al.}(1990)\citenamefont {Barger},
  \citenamefont {Hewett},\ and\ \citenamefont {Phillips}}]{Barger:1989}%
  \BibitemOpen
  \bibfield  {author} {\bibinfo {author} {\bibfnamefont {V.~D.}\ \bibnamefont
  {Barger}}, \bibinfo {author} {\bibfnamefont {J.}~\bibnamefont {Hewett}}, \
  and\ \bibinfo {author} {\bibfnamefont {R.}~\bibnamefont {Phillips}},\ }\href
  {\doibase 10.1103/PhysRevD.41.3421} {\bibfield  {journal} {\bibinfo
  {journal} {Phys.Rev.}\ }\textbf {\bibinfo {volume} {D41}},\ \bibinfo {pages}
  {3421} (\bibinfo {year} {1990})}\BibitemShut {NoStop}%
\bibitem [{\citenamefont {Gustafsson}\ \emph {et~al.}(2007)\citenamefont
  {Gustafsson}, \citenamefont {Lundstr\"om}, \citenamefont {Bergstr\"om},\ and\
  \citenamefont {Edsj\"o}}]{Gustafsson:2007}%
  \BibitemOpen
  \bibfield  {author} {\bibinfo {author} {\bibfnamefont {M.}~\bibnamefont
  {Gustafsson}}, \bibinfo {author} {\bibfnamefont {E.}~\bibnamefont
  {Lundstr\"om}}, \bibinfo {author} {\bibfnamefont {L.}~\bibnamefont
  {Bergstr\"om}}, \ and\ \bibinfo {author} {\bibfnamefont {J.}~\bibnamefont
  {Edsj\"o}},\ }\href {\doibase 10.1103/PhysRevLett.99.041301} {\bibfield
  {journal} {\bibinfo  {journal} {Phys. Rev. Lett.}\ }\textbf {\bibinfo
  {volume} {99}},\ \bibinfo {pages} {041301} (\bibinfo {year}
  {2007})}\BibitemShut {NoStop}%
\bibitem [{\citenamefont {Lopez~Honorez}\ \emph {et~al.}(2007)\citenamefont
  {Lopez~Honorez}, \citenamefont {Nezri}, \citenamefont {Oliver},\ and\
  \citenamefont {Tytgat}}]{LopezHonorez:2006}%
  \BibitemOpen
  \bibfield  {author} {\bibinfo {author} {\bibfnamefont {L.}~\bibnamefont
  {Lopez~Honorez}}, \bibinfo {author} {\bibfnamefont {E.}~\bibnamefont
  {Nezri}}, \bibinfo {author} {\bibfnamefont {J.~F.}\ \bibnamefont {Oliver}}, \
  and\ \bibinfo {author} {\bibfnamefont {M.~H.}\ \bibnamefont {Tytgat}},\
  }\href {\doibase 10.1088/1475-7516/2007/02/028} {\bibfield  {journal}
  {\bibinfo  {journal} {JCAP}\ }\textbf {\bibinfo {volume} {0702}},\ \bibinfo
  {pages} {028} (\bibinfo {year} {2007})},\ \Eprint
  {http://arxiv.org/abs/hep-ph/0612275} {arXiv:hep-ph/0612275 [hep-ph]}
  \BibitemShut {NoStop}%
\bibitem [{\citenamefont {Lopez~Honorez}(2007)}]{LopezHonorez:2007}%
  \BibitemOpen
  \bibfield  {author} {\bibinfo {author} {\bibfnamefont {L.}~\bibnamefont
  {Lopez~Honorez}},\ }\href@noop {} {\  (\bibinfo {year} {2007})},\ \Eprint
  {http://arxiv.org/abs/0706.0186} {arXiv:0706.0186 [hep-ph]} \BibitemShut
  {NoStop}%
\bibitem [{\citenamefont {Tytgat}(2008)}]{Tytgat:2007}%
  \BibitemOpen
  \bibfield  {author} {\bibinfo {author} {\bibfnamefont {M.~H.}\ \bibnamefont
  {Tytgat}},\ }\href {\doibase 10.1088/1742-6596/120/4/042026} {\bibfield
  {journal} {\bibinfo  {journal} {J.Phys.Conf.Ser.}\ }\textbf {\bibinfo
  {volume} {120}},\ \bibinfo {pages} {042026} (\bibinfo {year} {2008})},\
  \Eprint {http://arxiv.org/abs/0712.4206} {arXiv:0712.4206 [hep-ph]}
  \BibitemShut {NoStop}%
\bibitem [{\citenamefont {Dolle}\ \emph {et~al.}(2010)\citenamefont {Dolle},
  \citenamefont {Miao}, \citenamefont {Su},\ and\ \citenamefont
  {Thomas}}]{Dolle-Miao:2009}%
  \BibitemOpen
  \bibfield  {author} {\bibinfo {author} {\bibfnamefont {E.}~\bibnamefont
  {Dolle}}, \bibinfo {author} {\bibfnamefont {X.}~\bibnamefont {Miao}},
  \bibinfo {author} {\bibfnamefont {S.}~\bibnamefont {Su}}, \ and\ \bibinfo
  {author} {\bibfnamefont {B.}~\bibnamefont {Thomas}},\ }\href {\doibase
  10.1103/PhysRevD.81.035003} {\bibfield  {journal} {\bibinfo  {journal}
  {Phys.Rev.}\ }\textbf {\bibinfo {volume} {D81}},\ \bibinfo {pages} {035003}
  (\bibinfo {year} {2010})},\ \Eprint {http://arxiv.org/abs/0909.3094}
  {arXiv:0909.3094 [hep-ph]} \BibitemShut {NoStop}%
\bibitem [{\citenamefont {Arina}\ \emph {et~al.}(2009)\citenamefont {Arina},
  \citenamefont {Ling},\ and\ \citenamefont {Tytgat}}]{Arina:2009}%
  \BibitemOpen
  \bibfield  {author} {\bibinfo {author} {\bibfnamefont {C.}~\bibnamefont
  {Arina}}, \bibinfo {author} {\bibfnamefont {F.-S.}\ \bibnamefont {Ling}}, \
  and\ \bibinfo {author} {\bibfnamefont {M.~H.}\ \bibnamefont {Tytgat}},\
  }\href {\doibase 10.1088/1475-7516/2009/10/018} {\bibfield  {journal}
  {\bibinfo  {journal} {JCAP}\ }\textbf {\bibinfo {volume} {0910}},\ \bibinfo
  {pages} {018} (\bibinfo {year} {2009})},\ \Eprint
  {http://arxiv.org/abs/0907.0430} {arXiv:0907.0430 [hep-ph]} \BibitemShut
  {NoStop}%
\bibitem [{\citenamefont {Lopez~Honorez}\ and\ \citenamefont
  {Yaguna}(2010)}]{Honorez:2010}%
  \BibitemOpen
  \bibfield  {author} {\bibinfo {author} {\bibfnamefont {L.}~\bibnamefont
  {Lopez~Honorez}}\ and\ \bibinfo {author} {\bibfnamefont {C.~E.}\ \bibnamefont
  {Yaguna}},\ }\href {\doibase 10.1007/JHEP09(2010)046} {\bibfield  {journal}
  {\bibinfo  {journal} {JHEP}\ }\textbf {\bibinfo {volume} {1009}},\ \bibinfo
  {pages} {046} (\bibinfo {year} {2010})},\ \Eprint
  {http://arxiv.org/abs/1003.3125} {arXiv:1003.3125 [hep-ph]} \BibitemShut
  {NoStop}%
\bibitem [{\citenamefont {Lopez~Honorez}\ and\ \citenamefont
  {Yaguna}(2011)}]{LopezHonorez:2010}%
  \BibitemOpen
  \bibfield  {author} {\bibinfo {author} {\bibfnamefont {L.}~\bibnamefont
  {Lopez~Honorez}}\ and\ \bibinfo {author} {\bibfnamefont {C.~E.}\ \bibnamefont
  {Yaguna}},\ }\href {\doibase 10.1088/1475-7516/2011/01/002} {\bibfield
  {journal} {\bibinfo  {journal} {JCAP}\ }\textbf {\bibinfo {volume} {1101}},\
  \bibinfo {pages} {002} (\bibinfo {year} {2011})},\ \Eprint
  {http://arxiv.org/abs/1011.1411} {arXiv:1011.1411 [hep-ph]} \BibitemShut
  {NoStop}%
\bibitem [{\citenamefont {Sokolowska}(2011{\natexlab{a}})}]{Sokolowska:2011}%
  \BibitemOpen
  \bibfield  {author} {\bibinfo {author} {\bibfnamefont {D.}~\bibnamefont
  {Sokolowska}},\ }\href@noop {} {\  (\bibinfo {year} {2011}{\natexlab{a}})},\
  \Eprint {http://arxiv.org/abs/1107.1991} {arXiv:1107.1991 [hep-ph]}
  \BibitemShut {NoStop}%
\bibitem [{\citenamefont
  {Sokolowska}(2011{\natexlab{b}})}]{Sokolowska:2011-acta}%
  \BibitemOpen
  \bibfield  {author} {\bibinfo {author} {\bibfnamefont {D.}~\bibnamefont
  {Sokolowska}},\ }\href {\doibase 10.5506/APhysPolB.42.2237} {\bibfield
  {journal} {\bibinfo  {journal} {Acta Phys.Polon.}\ }\textbf {\bibinfo
  {volume} {B42}},\ \bibinfo {pages} {2237} (\bibinfo {year}
  {2011}{\natexlab{b}})},\ \Eprint {http://arxiv.org/abs/1112.2953}
  {arXiv:1112.2953 [hep-ph]} \BibitemShut {NoStop}%
\bibitem [{\citenamefont {Gorczyca}\ and\ \citenamefont
  {Krawczyk}(2011{\natexlab{b}})}]{Gorczyca:2011kr}%
  \BibitemOpen
  \bibfield  {author} {\bibinfo {author} {\bibfnamefont {B.}~\bibnamefont
  {Gorczyca}}\ and\ \bibinfo {author} {\bibfnamefont {M.}~\bibnamefont
  {Krawczyk}},\ }\href@noop {} {\  (\bibinfo {year} {2011}{\natexlab{b}})},\
  \Eprint {http://arxiv.org/abs/1112.5086} {arXiv:1112.5086 [hep-ph]}
  \BibitemShut {NoStop}%
\bibitem [{\citenamefont {Barate}\ \emph {et~al.}(2003)\citenamefont {Barate}
  \emph {et~al.}}]{LEP-modelind}%
  \BibitemOpen
  \bibfield  {author} {\bibinfo {author} {\bibfnamefont {R.}~\bibnamefont
  {Barate}} \emph {et~al.} (\bibinfo {collaboration} {LEP Working Group for
  Higgs boson searches, ALEPH Collaboration, DELPHI Collaboration, L3
  Collaboration, OPAL Collaboration}),\ }\href {\doibase
  10.1016/S0370-2693(03)00614-2} {\bibfield  {journal} {\bibinfo  {journal}
  {Phys.Lett.}\ }\textbf {\bibinfo {volume} {B565}},\ \bibinfo {pages} {61}
  (\bibinfo {year} {2003})},\ \Eprint {http://arxiv.org/abs/hep-ex/0306033}
  {arXiv:hep-ex/0306033 [hep-ex]} \BibitemShut {NoStop}%
\bibitem [{\citenamefont {Guedes}\ \emph {et~al.}(2012)\citenamefont {Guedes},
  \citenamefont {Moretti},\ and\ \citenamefont {Santos}}]{Guedes:2012}%
  \BibitemOpen
  \bibfield  {author} {\bibinfo {author} {\bibfnamefont {R.}~\bibnamefont
  {Guedes}}, \bibinfo {author} {\bibfnamefont {S.}~\bibnamefont {Moretti}}, \
  and\ \bibinfo {author} {\bibfnamefont {R.}~\bibnamefont {Santos}},\ }\href
  {\doibase 10.1007/JHEP10(2012)119} {\bibfield  {journal} {\bibinfo  {journal}
  {JHEP}\ }\textbf {\bibinfo {volume} {1210}},\ \bibinfo {pages} {119}
  (\bibinfo {year} {2012})},\ \Eprint {http://arxiv.org/abs/1207.4071}
  {arXiv:1207.4071 [hep-ph]} \BibitemShut {NoStop}%
\bibitem [{\citenamefont {Eriksson}\ \emph {et~al.}(2010)\citenamefont
  {Eriksson}, \citenamefont {Rathsman},\ and\ \citenamefont
  {Stal}}]{Eriksson:2010}%
  \BibitemOpen
  \bibfield  {author} {\bibinfo {author} {\bibfnamefont {D.}~\bibnamefont
  {Eriksson}}, \bibinfo {author} {\bibfnamefont {J.}~\bibnamefont {Rathsman}},
  \ and\ \bibinfo {author} {\bibfnamefont {O.}~\bibnamefont {Stal}},\ }\href
  {\doibase 10.1016/j.cpc.2009.12.016} {\bibfield  {journal} {\bibinfo
  {journal} {Comput.Phys.Commun.}\ }\textbf {\bibinfo {volume} {181}},\
  \bibinfo {pages} {833} (\bibinfo {year} {2010})}\BibitemShut {NoStop}%
\end{thebibliography}%

\end{document}